\documentclass[
  journal=pasa,
  manuscript=research-paper, 
  year=202X,
  volume=YY,
]{cup-journal}

\usepackage{bm}
\usepackage{amsmath}
\usepackage{amssymb,microtype,siunitx,booktabs}
\newcommand{\ha}{H$_{\alpha}$}
\newcommand{\EWha}{${\rm EW}_{{\rm H}\alpha}$}

\title{Exploring the spatially resolved initial mass function in SAMI star-forming galaxies}

\author{Diego Salvador}
\affiliation{School of Mathematical and Physical Sciences, 12 Wally’s Walk, Macquarie University, NSW 2109, Australia.}
\email[D. Salvador]{diego.salvador@hdr.mq.edu.au}

\author{Andrew M. Hopkins}
\affiliation{School of Mathematical and Physical Sciences, 12 Wally’s Walk, Macquarie University, NSW 2109, Australia.}

\author{Matt Owers}
\affiliation{School of Mathematical and Physical Sciences, 12 Wally’s Walk, Macquarie University, NSW 2109, Australia.}
\alsoaffiliation{ARC Centre of Excellence for All Sky Astrophysics in 3 Dimensions (ASTRO 3D), Australia.}

\author{Themiya Nanayakkara}
\affiliation{Centre for Astrophysics and Supercomputing, Swinburne University of Technology, P.O. Box 218, Hawthorn, 3122, VIC, Australia.}

\author{Scott M. Croom}
\affiliation{ARC Centre of Excellence for All Sky Astrophysics in 3 Dimensions (ASTRO 3D), Australia.}
\alsoaffiliation{Sydney Institute for Astronomy, School of Physics, University of Sydney, NSW 2006, Australia.}

\received {dd Mmm YYYY}
\revised  {dd Mmm YYYY}
\accepted {dd Mmm YYYY}
\published{DD MMMMMMMMM 2024}


\begin{document}

\begin{abstract}
The initial mass function (IMF) is a construct that describes the distribution of stellar masses for a newly formed population of stars. It is a fundamental element underlying all of star and galaxy formation, and has been the subject of extensive investigation for more than 60 years. In the past few decades there has been a growing, and now substantial, body of evidence supporting the need for a variable IMF. In this light, it is crucial to investigate the IMF's characteristics across different spatial scales and to understand the factors driving its variability. We make use of spatially resolved spectroscopy to examine the high-mass IMF slope of star-forming galaxies within the SAMI survey. By applying the Kennicutt method and stellar population synthesis models, we estimated both the spaxel-resolved ($\alpha_{res}$) and galaxy-integrated ($\alpha_{int}$) high-mass IMF slopes of these galaxies. Our findings indicate that the resolved and integrated IMF slopes exhibit a near 1:1 relationship for $\alpha_{int}\gtrsim -2.7$. We observe a wide range of $\alpha_{res}$ distributions within galaxies. To explore the sources of this variability, we analyse the relationships between the resolved and integrated IMF slopes and both the star formation rate (SFR) and SFR surface density ($\Sigma_{\rm{SFR}}$). Our results reveal a strong correlation where flatter/steeper slopes are associated with higher/lower SFR and $\Sigma_{\rm{SFR}}$. This trend is qualitatively similar for resolved and global scales. Additionally, we identify a mass dependency in the relationship with SFR, though none was found in the relation between the resolved slope and $\Sigma_{\rm{SFR}}$. These findings suggest an scenario where the formation of high-mass stars is favoured in regions with more concentrated star formation. This may be a consequence of the reduced fragmentation of molecular clouds, which nonetheless accrete more material. 
\end{abstract}

\section{INTRODUCTION}
\label{sec:int}

Galaxy evolution arises from the complex interactions between gravitational, hydrodynamical, and radiative processes \citep[e.g][]{somerville_2015}. The evolutionary mechanisms governed by these processes are, to a large extent, dependent on the distribution of stellar masses \citep{hopkins_2018}. This ``initial mass function'' (IMF) characterises this distribution within a given volume following a star formation event. As mass is the primary determinant of a star's evolutionary path, the IMF is crucial for comprehending star formation \citep{kennicutt_1998,hoversten_2008,elmegreen_2009,lee_2009,meurer_2009,gunawardhana_2011}, chemical enrichment \citep{portinari_1997,tornatore_2007,komiya_2011,goswami_2021}, feedback processes \citep{dib_2010,smith_2021}, mass-to-light ratios \citep{portinari_2004,cappellari_2012,mehrgan_2024}, and dark matter content \citep{rocca_1990,treu_2010}.

The IMF is commonly built as a piecewise power law, with each segment defined by a number $\alpha$ that denotes the slope of the power law within that mass range. This parameter reflects the distribution of stars such that $dN/dM \propto m^\alpha$ \citep{salpeter_1955,kroupa_1993,kroupa_2001}. Treating the IMF as an observable is particularly challenging as most massive stars ($\gtrsim 30 M\odot$) have typically evolved off the main sequence within 1 Myr, while lower-mass stars ($\lesssim  30 M\odot$) still remain in the main sequence \citep{Kroupa+2013}. Consequently, it is not possible to capture the complete mass distribution at a single point in time. Instead, the IMF is commonly derived by observing and analysing several stellar populations across various stages of evolution, allowing for a statistical estimation of the overall distribution \citep{kroupa_2001,Kroupa+2013,chabrier_2014,hopkins_2018,Li_2023}.

Given the statistical nature of the IMF, several methods have been developed to estimate it. These include the Kennicutt method \citep{kennicutt_1983,kennicutt_1994} for star forming galaxies, which uses the H$\alpha$ line equivalent width (\EWha) and an optical colour (such as $B-V$ or $g-r$) to derive the IMF. Other techniques involve dynamical approaches, such as gravitational lensing \citep{ferreras_2010,treu_2010,leier_2016} and velocity dispersion in early-type galaxies \citep{ferreras_2012,zaritsky_2012,esdaile_2021}. Additionally, spectral features are also employed to infer the IMF, including the $8183-8195$\,\AA\ Na I doublet \citep{vandokkum_2010,smith_2015}, the $9916$\,\AA\ Wing-Ford FeH molecular band \citep{hardy_1990,vandokkum_2010,conroy_2012,labarbera_2016,vaughan_2018}, the $8600\,\mu$m Ca II triplet \citep{saglia_2002,vazdekis_2003} and the TiO molecular band \citep{navarro_2015,labarbera_2016,labarbera_2017} for non-star forming galaxies, and the $2300$\,\AA\ CO index \citep{mieske_2008} and the $^{13}$C/$^{18}$O ratio \citep{zhang_2018,Brown_2019} for star forming galaxies.

It is commonly assumed, either for convenience or lack of strong evidence otherwise, that the IMF is universal \citep{scalo_1986,kroupa_2001,bastian_2010,hopkins_2013}, but there is strong conjecture around that point \citep{kennicutt_1998,hoversten_2008,meurer_2009,hoversten_2010,treu_2010,vandokkum_2010,gunawardhana_2011,narayanan_2012,nanayakkara_2017,smith_2014}. A Universal IMF is supported by the observation that star formation occurs in a variety of settings with differing densities and chemical compositions, yet most IMF estimations converge on a slope of $\alpha = -2.35$, known as the Salpeter slope. However, this notion has been debated since the early 1960s, with arguments supporting a varying IMF \citep{schmidt_1963}. More recent studies have highlighted the necessity for varying IMFs to account for differences in metallicity \citep{navarro_2015b,Zonoozi_16}, [Mg/Fe] content \citep{vandokkum_2010,conroy_2012}, SFR \citep{gunawardhana_2011}, radial distance \citep{navarro_2015}, velocity dispersion \citep{ferreras_2012,zaritsky_2012,conroy_2013,pernet_2024}, $F_{\text{H}\alpha}/F_{UV}$ flux ratio \citep{meurer_2009}, surface brightness \citep{hoversten_2010}, stellar mass-to-light ratios \citep{cappellari_2012}, stellar orbits \citep{porci_2022}, environment \citep{geha_2013,chabrier_2014} and redshift \citep{nanayakkara_2017}. Moreover, a variable IMF could be invoked to explain phenomena such as the G-dwarf problem \citep{worthey_1996} and the thickness and tilt of the fundamental plane of elliptical galaxies \citep{graves_2010}.

Several approaches for quantifying IMF shapes in star forming galaxies are now well established \citep{kennicutt_1983,kennicutt_1994,buat_1987,meurer_2009}, and models such as the IGIMF \citep{kroupa_2003} have been developed to explore variations found using these approaches. Given the advent of large scale resolved spectroscopy datasets, such as SAMI \citep{croom_2012}, CALIFA \citep{sanchez_2012}, and MaNGA \citep{bundy_2015}, it is now possible to apply such techniques to explore how the IMF may vary within galaxies. Notably, the work of \cite{navarro_2015,navarro_2019}, \cite{labarbera_2016,Labarbera_2019}, \cite{Parikh_2018} and \cite{barbosa_2021} stands out for their resolved IMF estimations using spectral features in early-type galaxies and \cite{navarro_2023} on the late-type galaxy NGC 3351. Most of these studies focus on low-mass stars in non-star-forming galaxies, thus reflecting an already established IMF.

In this work, we explore resolved IMF measurements within star forming galaxies using integral field spectroscopy (IFS) data from the SAMI survey DR3 \citep{croom_2021}. We perform spaxel-scale measurements of the high-mass IMF slope using the Kennicutt method \citep{kennicutt_1983,kennicutt_1994}. This is the first large-scale exploration of resolved IMF measurements in star forming galaxies. This approach provides a novel perspective on the IMF slope distribution within galaxies. It enables comparisons between local IMF slope estimations and those derived from the traditional integrated galaxy light. Because the SAMI sample is drawn in part from the Galaxy And Mass Assembly (GAMA) galaxy survey \citep{driver_2022} we can compare directly to analogous global IMF measurements for those galaxies \citep{gunawardhana_2011}. Our goal is to establish a new method for extracting new information about the variation of the IMF within galaxies, and trying to understand and link this information with the already-existing global galaxy integrated light IMF measurements. Additionally, we compare the resolved IMF measurements with other resolved physical quantities, such as star formation rate (SFR), star formation rate density ($\Sigma_{\rm SFR}$) and the galaxy stellar mass, to investigate potential underlying mechanisms driving the IMF shapes. These can again be compared against other published results \citep{lee_2009,meurer_2009,gunawardhana_2011,weidnet_2013}. Our analyses will enhance our understanding of the interplay between the IMF and various galactic properties, shedding light on the mechanisms driving galaxy evolution.

The layout of this paper is as follows. In \S\,\ref{sec:data} we describe the SAMI data products and the sample we use. In \S\,\ref{sec:methods} we describe the Kennicutt method and the population synthesis model required to make the IMF slope estimations. In \S\,\ref{sec:results} we describe our results. In \S\,\ref{sec:discussion} we address the limitations of this work, contextualises our results within the framework of existing research, and explores their implications. Finally, in \S\,\ref{sec:conclusion} we summarise our findings and key remarks.

Throughout we assume cosmological parameters of $H_0=70\,$km\,s$^{-1}$\,Mpc$^{-1}$, $\Omega_M=0.3$, $\Omega_\Lambda=0.7$ and $\Omega_{\rm{k}} = 0$.

\section{DATA}
\label{sec:data}

We use the data products provided by the SAMI Galaxy Survey Data Release 3 \citep{croom_2021}. The survey's data were collected using the Sydney-AAO Multi-object Integral field spectrograph (SAMI; \citealt{croom_2012}), located at the 3.9-metre Anglo-Australian Telescope (AAT) at Siding Spring Observatory. SAMI consists of 13 fibre bundles, called hexabundles, inserted into a custom field plate of 1-degree diameter on the sky at the corrected prime focus of the 3.9-metre aperture of the AAT. Each of these IFUs consists of 61 elements of 1.6 arcsec each, resulting in a field of view (FoV) of 15 arcsec for each IFU. This setup allows the instrument to perform multiplexed observations, with 12 IFUs observing galaxies and one calibration star simultaneously.

The SAMI Galaxy Survey \citep{croom_2012,bryant_2015,croom_2021},  was conducted between 2013 and 2018, collecting data on 3068 galaxies covering a broad range of stellar masses ($10^7-10^{12}M_\odot$) and environments (field, galaxy groups, and clusters) at $z<0.095$. Targets were primarily selected from 144 deg$^2$ equatorial regions from the GAMA survey \citep{driver_2009}, and 8 cluster regions described in \cite{owers_2017}.

The SAMI IFU allows for spatially resolved spectroscopy of each galaxy, providing spectra for several regions across the galaxy \citep{bryant_2015}. The primary SAMI data products are two data cubes, each formatted as a $2048 \times 50 \times 50$ array, with 2048 representing the wavelength axis and $50x50$ representing spatial coordinates. The wavelength ranges for the blue and red cubes are $3630-5800$\,\AA\ and $6250-7460$\,\AA, respectively. Each spaxel (spatial pixel) has a pixel scale of 0.5 arcseconds, with a spectral (wavelength) pixel scale of $1.04$,\AA\ per pixel for the blue cubes and $0.57$,\AA\ per pixel for the red cubes. The spectral resolutions are $R = 1808$ ($\sigma = 70.4$ km s$^{-1}$) for the blue arm and $R = 4304$ ($\sigma = 29.6$ km s$^{-1}$) for the red arm. The data reduction and cubing process is described elsewhere \citep{allen_2014,sharp_2015,green_2018,scott_2018,croom_2021}. The point spread function for each observation, with an average FWHM of 2.06 arcsec, is determined by fitting a Moffat profile to the flux distribution of the secondary standard star \citep{scott_2018}.

We also use other SAMI DR3 data products, such as two-dimensional maps of total emission line fluxes for H$\alpha$ and H$\beta$, gas velocity, gas velocity dispersion, star formation rate (SFR), star formation rate density ($\Sigma_{\rm SFR}$), and SFR masks for star-forming spaxels. Line flux maps are created by summing all the flux based on a single component Gaussian fit associated with a given emission line in each spaxel. \ha\ maps are extinction-corrected using the Balmer decrement and the \citep{cardelli_1989} extinction law. Gas velocity and velocity dispersion maps are obtained by fitting 11 strong optical emission lines with a single Gaussian component using LZIFU (LaZy-IFU, \citealt{ho_2016}). SFR maps (in $M_\odot \, \text{yr}^{-1}$) are derived from the extinction-corrected \ha\ maps, assuming $ \text{SFR} = L(\text{H}\alpha) \times \left(\frac{7.9}{1.53}\right) \times 10^{-42} [M_\odot \, \text{yr}^{-1}]$ \citep{medling_2018}. $\Sigma_{\rm SFR}$ (in $M_\odot \, \text{yr}^{-1} \, \text{kpc}^{-2}$) maps are provided similarly. Each spaxel emission is classified as star-forming, in a SFR mask, according to the BPT/VO87 diagnostics from \cite{kewley_2006}, based on \cite{kewley_2001} and \cite{kauffmann_2003} [OIII]/H$\beta$, [NII]/H$\alpha$, [SII]/H$\alpha$, and [OI]/H$\alpha$ flux ratio dividing lines.

We also use other physical quantities available in the SAMI DR3 tables. From \texttt{InputCatGAMADR3} and  \texttt{InputCatClustersDR3}, we extract stellar masses and effective radius, while H$\alpha_{Re}$ and H$\beta_{Re}$ fluxes, SFR$_{Re}$, and spectroscopic redshift are retrieved from \texttt{EmissionLine1compDR3}. Note that the $R_e$ sub-index indicates that these quantities are estimated from an elliptical aperture with a semi-major axis of 1 effective radius ($R_e$), where $R_e$ was measured using a single-Sérsic fit \citep{kelvin_2012, owers_2019}. We also use the blue and red 1 $R_e$ aperture spectra to calculate the galaxy integrated light colours and equivalent widths.

\begin{figure*}
\centering
\includegraphics[width=0.8\linewidth]{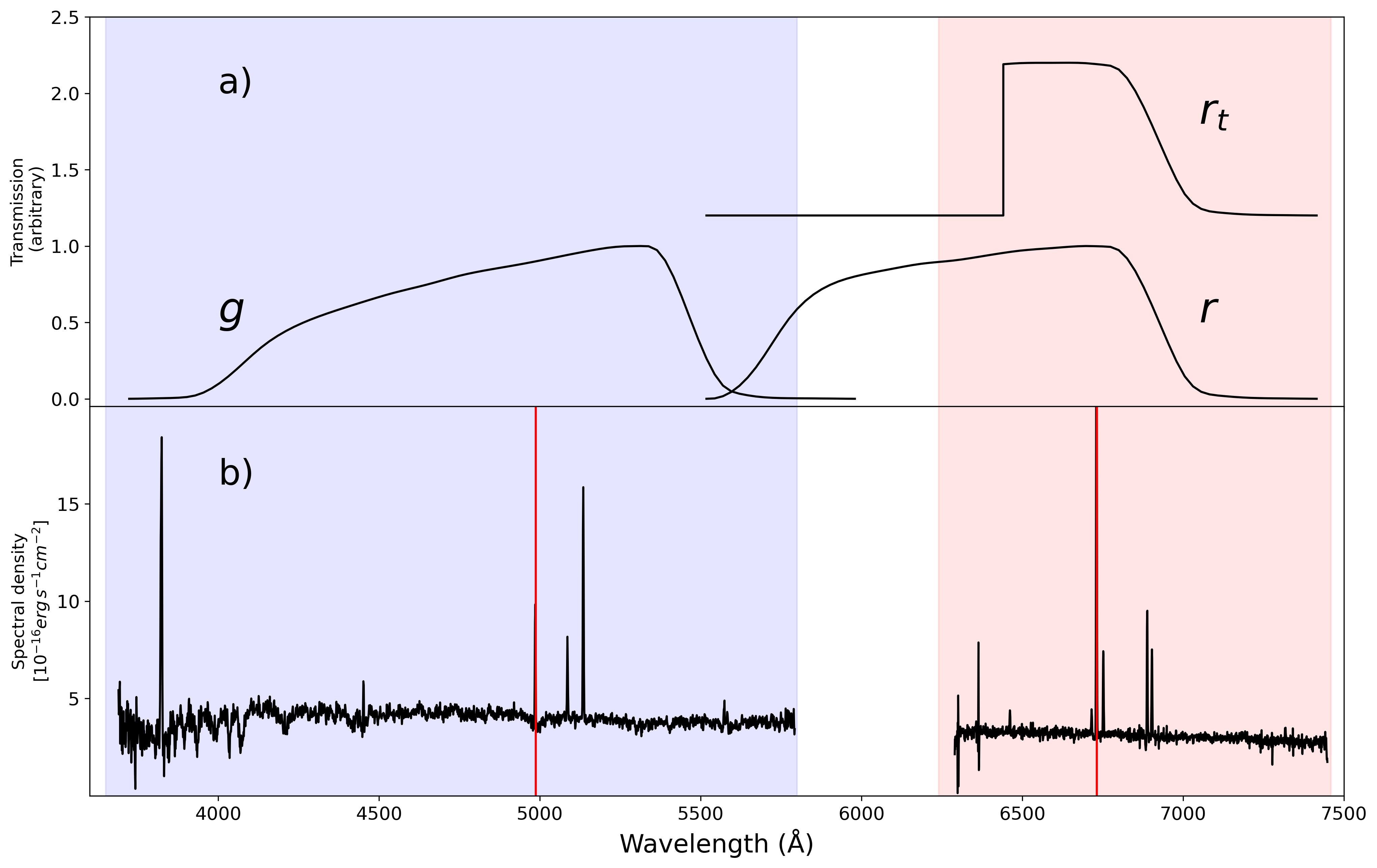}
\caption{\textit{a)} The normalised filter response functions for the $g$, $r$, and $r_t$ bands. \textit{b)} An example spectrum of a SAMI galaxy at $z=0.025$. The blue and red regions denote the wavelength ranges of the blue and red SAMI data cubes, respectively. The vertical red lines indicate the positions of the rest-frame wavelength of the H$\beta$ (left) and H$\alpha$ (right) emission lines. Bands have also been shifted to the rest-frame.}
\label{fig_spec}
\end{figure*}

As we aim to calculate the $g$- and $r$-band fluxes from the SAMI data using $g$ and $r$ SDSS filters, it is important to note that part of the bluest region of the $r$ band filter range ($5380-7230$\,\AA) falls outside the wavelength coverage of the SAMI red data cube ($6240-7460$\,\AA). To address this, we construct a truncated $r$ band, $r_t$ (Figure~\ref{fig_spec}), which covers the rest-frame wavelength range from $6240$\,\AA\ to 7460\,\AA. For consistency, we use this truncated band $r_t$ throughout the rest of the paper.

\begin{figure*}
\centering
\includegraphics[width=0.8\linewidth]{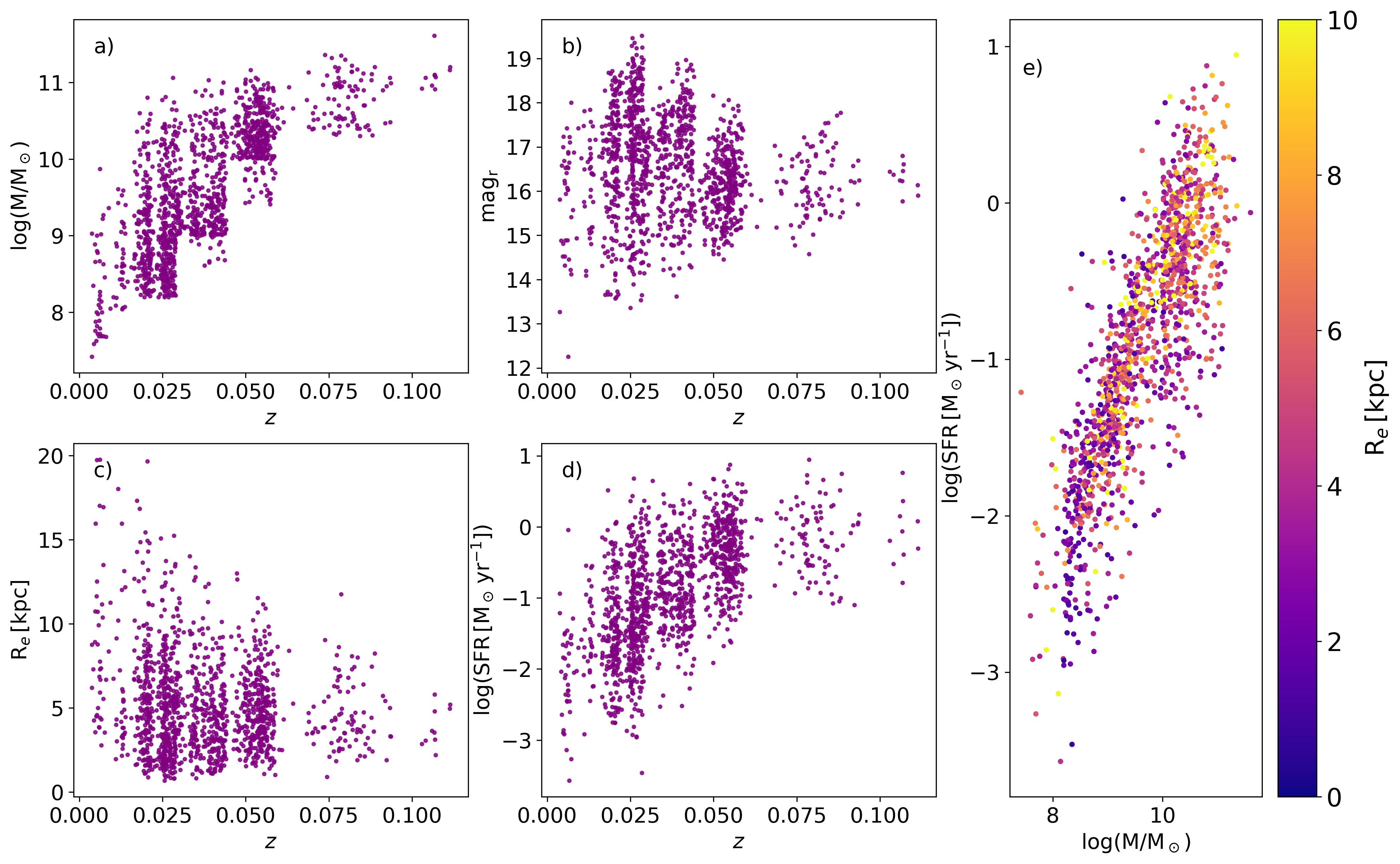}
\caption{This figure presents key properties of galaxies in our SAMI sample. Panel \textit{a)} illustrates the stellar mass distribution, while panel \textit{b)} shows the $r$-band magnitude, panel \textit{c)} highlights the effective radius of galaxies, and panel \textit{d)} displays the logarithmic star formation rate (SFR) as a function of redshift ($z$). Panel \textit{e)} maps the SFR-stellar mass relation, with points colour-coded by effective radius, demonstrating that the galaxies in our sample follow the typical star-forming main sequence. Our sample predominantly consists of high-stellar-mass galaxies due to the SAMI survey’s mass selection criteria, its focus on low redshift, and its exclusion of compact or very small systems.}
\label{data_exploration}
\end{figure*}

In this work we select all galaxies in the SAMI survey DR3 with $7 \leq log(M/M_\odot) \leq 12$ and a $10^{-3} \leq \rm{SFR} \leq 10^2$ within $0<z<1.12$. The Kennicutt method is generally unsuitable for passive galaxies due to their minimal or absent H$\alpha$ and UV emissions\footnote{Post-starburst galaxies may exhibit some \ha\ emission contribution from asymptotic giant branch (AGB) stars. However, given that this phase of stellar evolution is relatively short-lived, its impact on the overall \ha\ emission can be considered negligible.} from high-mass young stars \citep{kennicutt_1983}. Therefore, we restrict our analysis to galaxies identified as non-passive based on the spectroscopic classifications by \cite{owers_2019}. Additionally, our analysis exclusively considers spaxels classified as star-forming, as indicated by the SFR mask in the SAMI data products. The resulting sample contains 1344 objects. Figure~\ref{data_exploration} provides a visual summary of the physical properties of our sample.

\section{METHODS}
\label{sec:methods}

We make use of the Kennicutt diagnostic \citep{kennicutt_1983,kennicutt_1994,hoversten_2008,hoversten_2010,gunawardhana_2011,nanayakkara_2017} to estimate the IMF slope by comparing the \EWha line equivalent width with the $g-r$ colour. The \EWha\ is the ratio of the \ha\ flux, primarily contributed by gas ionisation by young massive stars ($>10M_\odot$), to the continuum at 6563\,\AA, sensitive to the older stellar populations. This ratio is, thus, sensitive to the high-mass star population. In contrast, the $g-r$ colour is sensitive to the low-mass star population, as cooler, low-mass stars emit less blue light than red light, resulting in a larger $g-r$ value. This makes the \EWha\ vs $g-r$ colour diagnostic an effective probe for the stellar IMF slope in star-forming galaxies.

We calculate \EWha\ as the ratio of the H$\alpha$ flux from the emission-line fits (LZIFU) to the continuum, which is estimated following the method outlined by \cite{cardiel_1998}. To correct the observed \EWha\ for obscuration, we follow the prescription in \cite{gunawardhana_2011}, which recovers the flux absorbed by dust by considering the gas and continuum colour excess derived from Balmer decrement (H$\alpha/$H$\beta$) and the obscuration curve from \cite{cardelli_1989}. We use Balmer decrement maps provided by SAMI, and thus we only consider this spaxels in our sample. It is important to note that the emission line fluxes in SAMI have already been corrected for underlying stellar absorption, so no additional correction is required. Colours are corrected for dust extinction using the reddening curve from \cite{calzetti_2001}.

We employ the \texttt{P{\'E}GASE}.3 population synthesis tool \citep{fioc_2019}, based on the original \texttt{P{\'E}GASE} software described in \cite{fioc_1997}, to construct galaxy evolutionary tracks in the \EWha\ vs $g-r_t$ space. \texttt{P{\'E}GASE} is a spectral synthesis code that models the spectral evolution and chemical composition of galaxy components. Given an input IMF form and specified low- and high-mass slopes, it generates galaxy spectra, from which it computes the normalised attenuated monochromatic continuum luminosity, in-band fluxes, and equivalent widths at various galactic ages. \texttt{P{\'E}GASE} models the \EWha\ emission by assuming that young, massive stars emit Lyman continuum photons, which ionise the surrounding gas. This results in nebular emission lines, including \EWha\ . Nebular emission is calculated using grids generated by the Cloudy code \citep{ferland_2017}. These models simulate dust-free HII regions as a function of ISM metallicity and the number of Lyman continuum photons emitted by the ionising source. The geometry used in these calculations is radiation-bounded and spherical, with constant hydrogen density throughout the ionised cloud. This allows us to create  a group of evolutionary tracks in the \EWha\ vs $g-r$ colour space, each of them associated with a different high-mass IMF slope. We have chosen \texttt{P{\'E}GASE} as it is one of the few stellar population synthesis (SPS) tools that allows to treat the IMF as a free parameter and modify its formalism, low and high mass range.

\begin{figure*}[hbt!]
\centering
\includegraphics[width=0.75\linewidth]{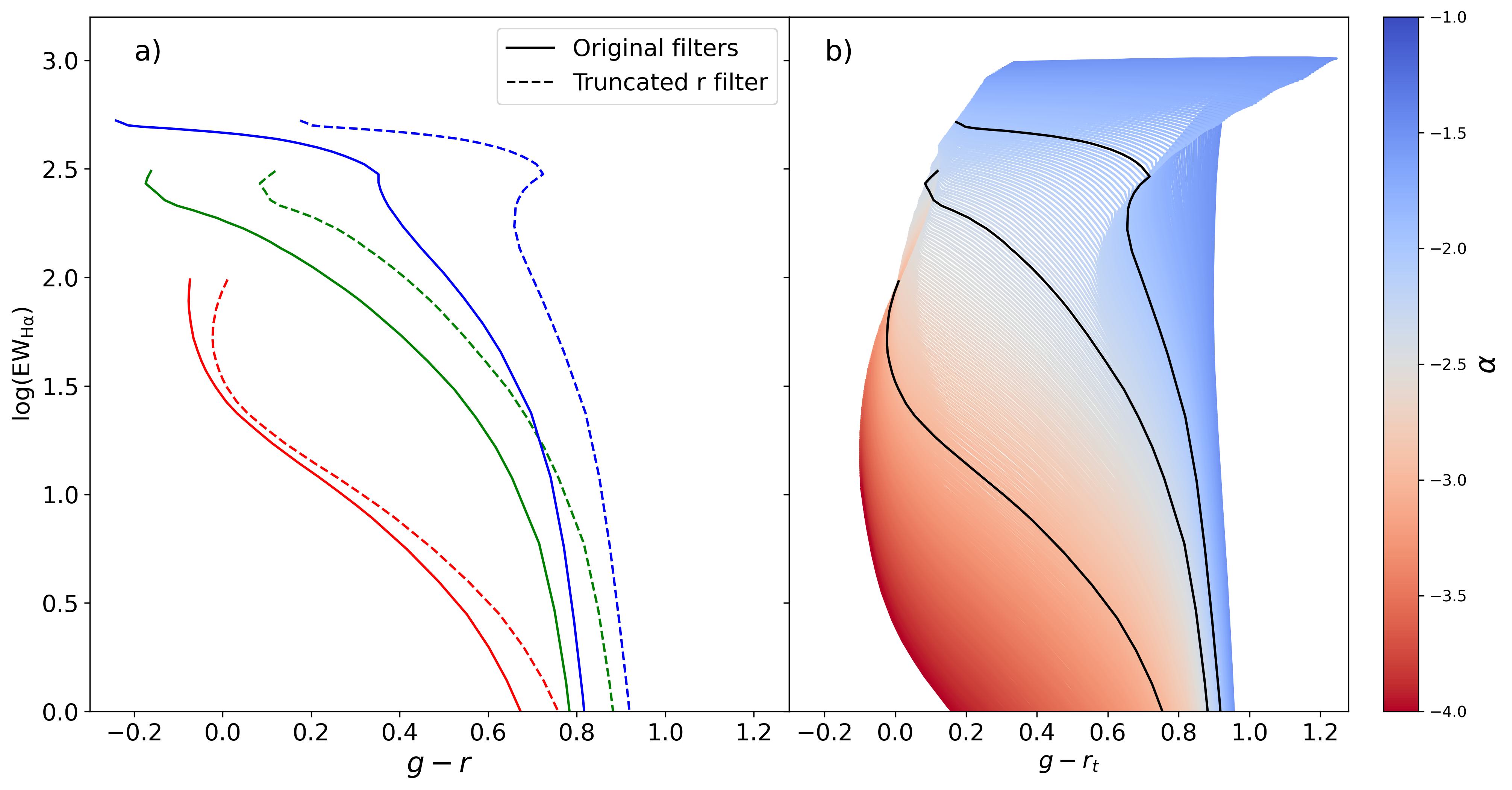}
\caption{\textit{a)} A comparison of the \texttt{P{\'E}GASE} evolutionary tracks generated using the original SDSS $g$ and $r$ band filters (dashed lines) with those generated using the truncated $r_t$ filter (solid lines). Each colour represents a track with a different IMF slope: red for $\alpha = -3.0$, green for $\alpha = -2.35$, and blue for $\alpha = -2.0$. The truncated $g-r_t$ colour is shifted to the right compared to the original $g-r$ colour due to the reduced flux in the $r_t$ band. \textit{b)} The $\log({\rm EW}_{{\rm H}\alpha})$ versus $g-r_t$ parameter space, populated by all generated \texttt{P{\'E}GASE} evolutionary tracks with the truncated filters. Each track corresponds to a different high-mass IMF slope, ranging from $\alpha=-1.5$ (top track) to $\alpha=-4.0$ (bottom track) in increments of 0.01.}
\label{tracks_comparison}
\end{figure*}

To adequately cover the range of IMF slopes in the \EWha\ vs $g-r$ colour space, we compute several evolutionary tracks using \texttt{P{\'E}GASE}. We consider a fixed low-mass slope of 0.5 and a variable high-mass slope ranging from  -4 to -1.5 in increments of 0.1. We consider the lower and upper stellar mass limits of our models to be 1$M_\odot$ and 120$M_\odot$, respectively. Other model assumptions include no infall, no extinction, no galactic winds, a non-evolving stellar metallicity of 0.02, and an exponentially declining star formation rate with an e-folding time of 1.1~Gyr. \texttt{P{\'E}GASE} calculates spectra at fixed ages ranging from 0 to 20~Gyr, with increasing time steps. However, we only consider ages greater than 100~Myr, as it takes approximately that time for \texttt{P{\'E}GASE} to stabilise the spectral energy distribution (SED) shape. All generated tracks are shown in panel \textit{b)} of Figure~\ref{tracks_comparison}. The location of the evolutionary tracks depends on several model input parameters, such as star formation histories, high-mass cutoffs and metallicities. It is not our objective to explore variations with these parameters, as they have been thoroughly investigated in prior works \citep{hoversten_2008,hoversten_2010,gunawardhana_2011,nanayakkara_2017,nanayakkara_2020}. Note that as we use a truncated $r$ colour, $r_t$, the evolutionary tracks are shifted towards larger (bluer) $g-r$ values, as the $r$ magnitudes are fainter due to less flux in the $r_t$ band compared to the full $r$. A comparison between three \texttt{P{\'E}GASE} tracks using the original $r$ band and the truncated $r_t$ band is depicted in the panel  \textit{a)} of Figure~\ref{tracks_comparison}.

The SAMI blue and red data cubes enable us to estimate the \EWha\ and the $g-r_t$ colour for all star-forming spaxels in every galaxy within our sample. We developed a method to assign an IMF slope to each spaxel based on its position in the $\log({\rm EW}_{{\rm H}\alpha})$ vs $g-r_t$ space. We use the evolutionary tracks from \texttt{P{\'E}GASE} and their associated IMF slopes to determine the IMF slope for each data point through an inverse distance weighted interpolation (IDWI) algorithm \citep{shepard_1968}. IDWI is a spatial interpolation method that interpolates between a group of points by considering the distance from known points to the point to be interpolated. Let $\bm{x_i}$ be a point in the parameter space, associated with a slope value $u_i$. Then, the interpolated value $u$ for an arbitrary point $\bm{x}$ is given by:
\begin{equation}
    u(\bm{x}) = \frac{\sum_{i=1}^N \omega_i (\bm{x})u_i}{\sum_{i=1}^N \omega_i (\bm{x})},
\end{equation}
where
\begin{equation}
    \omega_i=\frac{1}{d(\bm{x_i},\bm{x})^m},
\end{equation}
and $d(\bm{x_i},\bm{x})$ is the distance between the points $\bm{x}$ and $\bm{x_i}$. When $d(\bm{x_i},\bm{x})=0$, then $u(\bm{x})=u_i$. The power parameter $m$ typically ranges between 1 and 4. In this work we use $m=3$ as higher values of $m$ give greater importance to the closest values to the interpolated point.

\begin{figure*}[hbt!]
\centering
\includegraphics[width=0.8\linewidth]{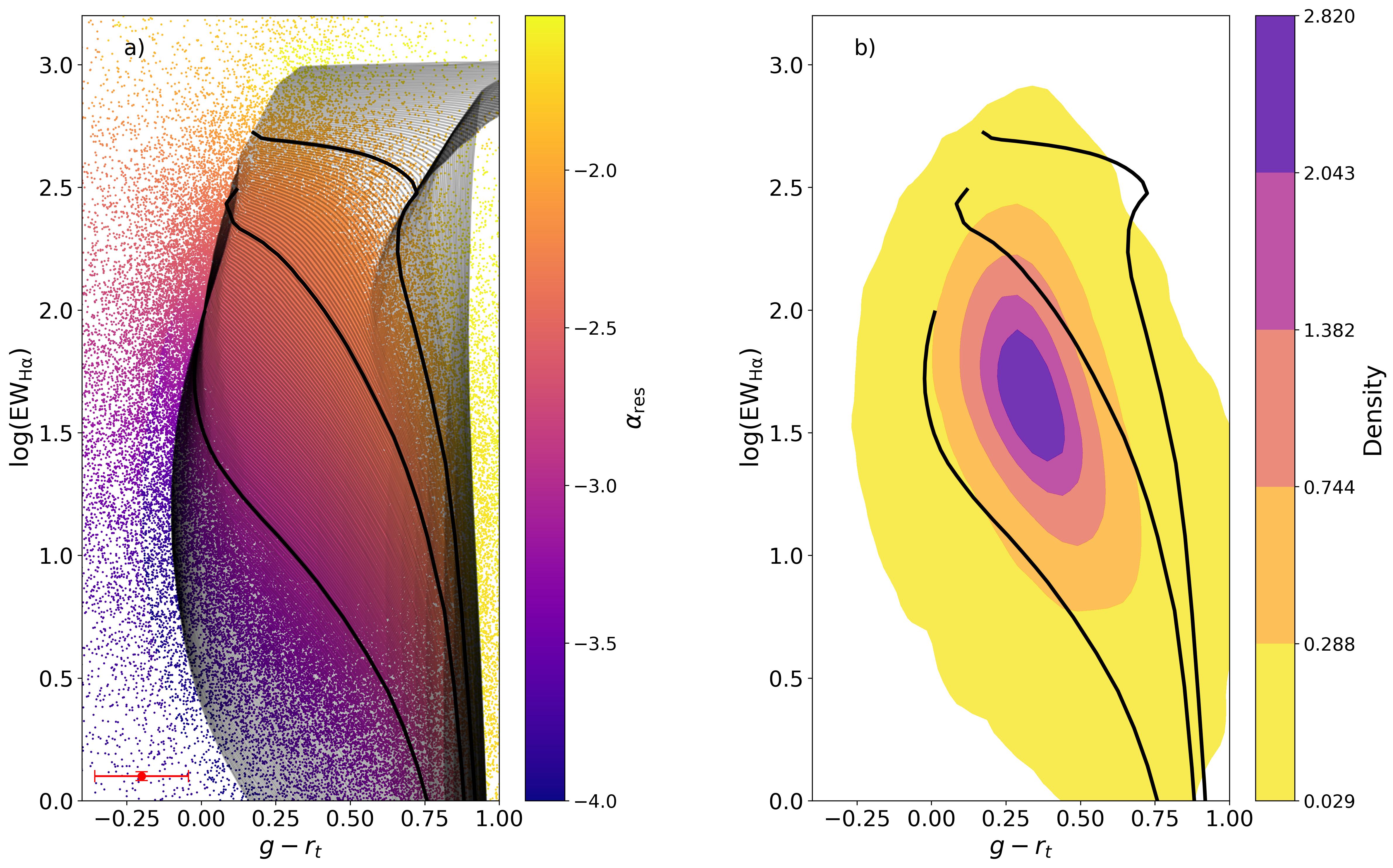}
\caption{Panel \textit{(a)} presents the $\log({\rm EW}_{{\rm H}\alpha})$ versus $g-r_t$ colour space, populated by spaxels from all galaxies in the sample. Each spaxel is colour-coded according to its high-mass IMF slope ($\alpha_{res}$), assigned through inverse distance weighted interpolation criterion. The black lines correspond to evolutionary tracks from the \texttt{PÉGASE} models, with each track representing a different high-mass IMF slope (flatter IMFs to the top-right). The thick black tracks highlight slopes of $-2$, $-2.35$, and $-3$ from top to bottom. A red cross in the bottom left corner indicates representative uncertainties along each axis. Panel \textit{(b)} shows the density contours of the spaxel distribution from panel \textit{(a)}, emphasising that the bulk of the data points lie within the region covered by the \texttt{PÉGASE} models. Notably, a significant portion of the data falls between the tracks associated with slopes of $-3$ and $-2.35$.}
\label{all_data_tracks}
\end{figure*}

In panel \textit{a)} of Figure~\ref{all_data_tracks} we show how all the spaxels of our sample distribute in the $\log({\rm EW}_{{\rm H}\alpha})$ vs $g-r_t$ space, colour-coded by their high-mass IMF slope. Our approach provides a novel perspective on the IMF distribution and variation in star-forming galaxies, highlighting that most data points cluster around $\alpha_{res}=-2$ and $\alpha_{res}=-2.35$ (panel \textit{b)} of Figure~\ref{all_data_tracks}), and that the spaxel distribution closely resembles that of galaxies (see Figure 4 in \citealt{gunawardhana_2011}). Notice that a group of spaxels fall outside the region enclosed by the outermost evolutionary tracks. For this set of spaxels our interpolation is inaccurate, as the IMF slope value should be extrapolated. However, only $\lesssim 8\%$ of the spaxels have this problem and thus our results should not be severely affected. 


To estimate the uncertainties of the IMF slopes, we use the $g-r_t$ colour uncertainty shown in the bottom left corner of Figure \ref{all_data_tracks}. We randomly select 1,000 points in the $\log({\rm EW}_{{\rm H}\alpha})$ versus $g-r_t$ plane, located within a rectangular box that contains the area with the highest density of points (regions with density greater than 0.288 in panel \textit{b} of Figure \ref{all_data_tracks}). For each point $(x,y)$, we apply the IDWI method to calculate the IMF slope at the coordinates $(x - err_{g-r_t}, y)$ and $(x + err_{g-r_t}, y)$, where $err_{g-r_t}$ is the colour error. We then find the difference between these two slope estimates. Since the uncertainty in \EWha\ is negligible and the models mainly vary along the $g-r_t$ axis, we do not consider variations in $EW_{\text{H}\alpha}$ when estimating uncertainty. The uncertainty in $\alpha_{\text{res}}$ is determined as the standard deviation of the differences calculated from the 1,000 points, yielding an uncertainty of $\Delta \alpha_{res} \sim 0.167$.

Since SAMI spaxels cover a spatial scale ranging from $\sim$ 70 to $\sim$ 700 parsecs, they encompass a larger area than a typical individual star-forming region. Consequently, concerns regarding the sampling of the initial mass function (IMF) are minimised.

\begin{figure}[hbt!]
\centering
\includegraphics[width=0.75\linewidth]{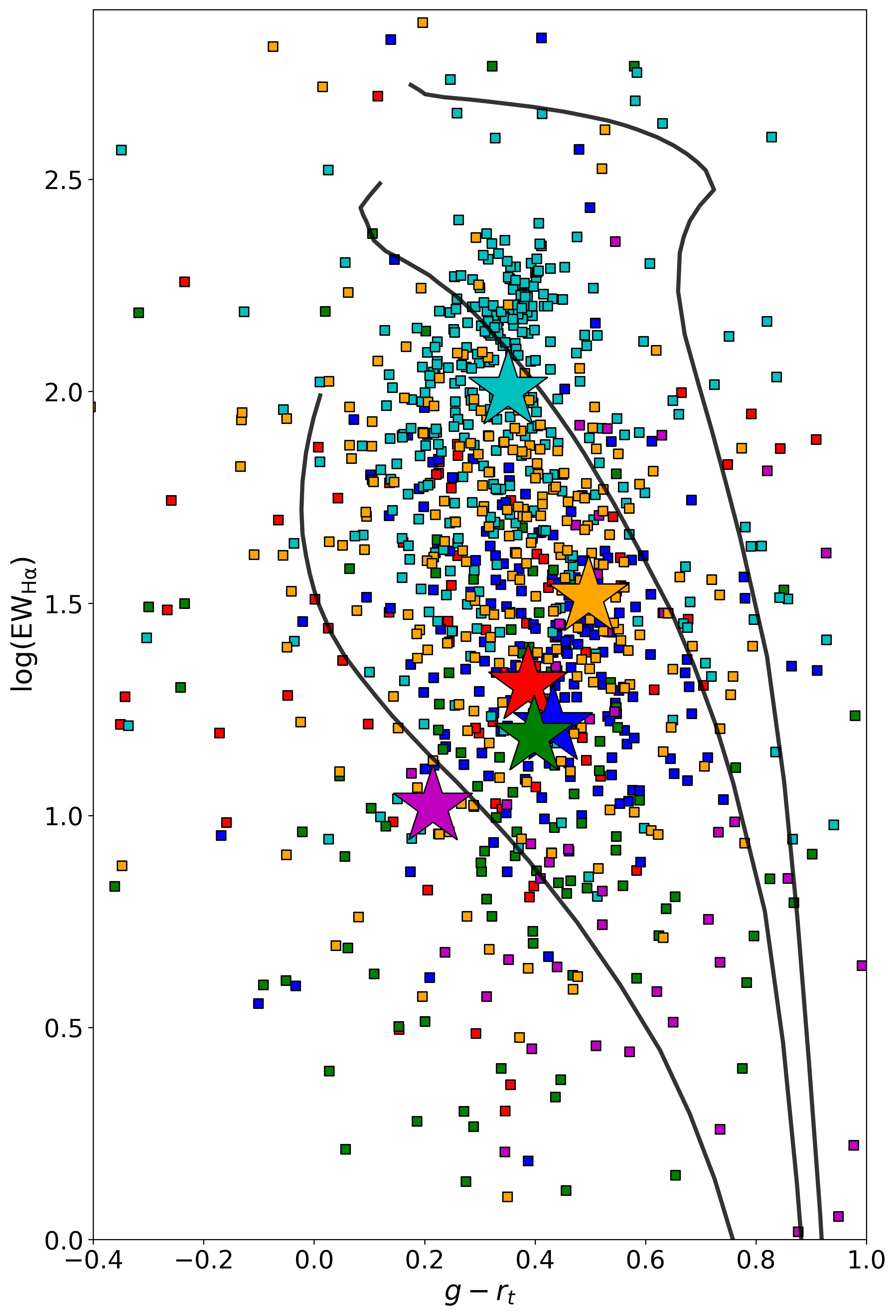}
\caption{The distribution of spaxels from six galaxies in our sample within the $\log({\rm EW}_{{\rm H}\alpha})$ versus $g-r_t$ colour parameter space. Spaxels from the same galaxy are represented as squares of the same colour, while the integrated galaxy measurements of $\log({\rm EW}_{{\rm H}\alpha})$ and $g-r_t$ (obtained using the $1-R_e$ aperture spectra) are shown as stars. The black lines correspond to three \texttt{P{\'E}GASE} tracks with $\alpha=-2$ (top), $\alpha=-2.35$ (middle) and $\alpha=-3$ (bottom).}
\label{all_data_tracks_coloured}
\end{figure}

We also apply the same procedure to the galaxy-integrated light measurements. Using the 1$R_e$ aperture SAMI red and blue cubes, we estimate the \EWha\ equivalent widths and $g-r_t$ colours, and use the same IDWI criterion to assign an IMF slope to each galaxy. We compare the distribution of spaxels versus the integrated measurements in the $\log({\rm EW}_{{\rm H}\alpha})$ vs $g-r_t$ space in Figure~\ref{all_data_tracks_coloured}. It is noteworthy that the spaxels within the same galaxy do not necessarily cluster around the integrated measurements. Additionally, both spaxels and integrated measurements exhibit a range of IMF slope values, both flatter and steeper than the Salpeter slope. These observations suggest that the IMF slope of a galaxy and the locally resolved IMF slope within that galaxy are not necessarily in a one-to-one relation. We explore the variability in $\alpha_{res}$ further in the next section.

We denote the IMF slope estimated for individual resolved spaxels as $\alpha_{res}$, while the IMF slope derived from global or integrated galaxy light measurements from the 1$R_e$ spectra is referred to as $\alpha_{int}$. In future sections, $\bar{\alpha}$ will refer to the IMF slope corresponding to the coordinates of maximum spaxel density in the $\log({\rm EW}_{{\rm H}\alpha})$ vs $g-r_t$ space.

\section{RESULTS}
\label{sec:results}

Understanding the spatial distribution of the IMF slope within galaxies is essential for grasping star formation processes and galaxy evolution. The IMF slope provides insights into the stellar mass distribution, reflecting variations in star formation efficiency and the star formation history of a galaxy. By examining how the IMF slope varies across different regions of a galaxy, we can gain a deeper understanding of the underlying physical conditions and processes influencing star formation.

To address these questions, we have produced resolved IMF slope maps for each galaxy in our sample using the SAMI IFU and the IMF slope estimation method described in \S\,\ref{sec:methods}. A selection of these maps is presented in Figure~\ref{IMF_map_9}, along with the spatial distribution maps of other quantities. The IMF slope maps exhibit diverse spatial distributions and morphologies. For example, galaxy 271562 exhibits a radial gradient where the IMF slope is flatter (more positive) in the centre and steeper (more negative) toward the periphery. In contrast, galaxy 220515 shows the opposite trend, with a steeper IMF slope at the centre and a flatter slope toward the edges. Notably, their SFR maps differ: 271562 shows no radial gradient, whereas 220515’s SFR map closely mirrors the gradient seen in its IMF slope. This underscores a relation between $\alpha_{res}$ and SFR which will be explored further below. Additionally, galaxy 517868 contains a localised region with steeper IMF slopes compared to the rest of the galaxy. In contrast, the maps for galaxy 618116 are relatively flat, with no clear patterns or distinct features. Lastly, galaxy 376478 lacks central star formation and exhibits a relatively flat IMF slope map with values around $\alpha_{res} \sim -2.6$. Notably, both galaxies 618116 and 376478 have very high SFRs but display steep $\alpha_{res}$ slopes. This suggests that when star formation is widely dispersed, the available gas is converted predominantly into numerous low-mass stars rather than a few high-mass stars, leading to more negative $\alpha_{res}$ values. The features observed in these maps of the $\alpha_{res}$ distribution are consistently reflected in the SFR, $g-r_t$, and \EWha\ maps.

Overall, the IMF slope spatial distribution appears neither uniform (i.e., consistent across the galaxy, typical standard deviation of $\sigma=0.31$) nor random (i.e., lacking any ordered pattern or smooth transition). The underlying causes of these distinct IMF slope morphologies must be analysed on a case-by-case basis. 

In this section we explore the relationship between the resolved and integrated IMF slopes, as well as the link between the resolved and integrated IMF slopes and stellar mass, SFR, and $\Sigma_{\rm SFR}$.

\subsection{Links between $\alpha_{res}$ and $\alpha_{int}$}
\label{subsec:resolved_integrated}

A key innovation and advantage of our work is the capability to estimate both the global and the spatially resolved IMF-slope. This enables us to address important questions about the relationship between these two quantities. Specifically, we aim to determine whether they are related, whether they are consistent with each other, or whether one can provide context for interpreting the other.

To relate the integrated IMF-slope of each galaxy to its spatially resolved IMF-slopes, we use a representative value for the galaxy’s typical resolved IMF-slope. We have chosen to use the IMF-slope associated with the point of maximum density ($\bar{\alpha}$) in the distribution of spaxels in the $\log({\rm EW}_{{\rm H}\alpha})$ vs $g-r_t$ parameter space for each galaxy\footnote{In this approach, we calculate the maximum density coordinates of the spaxels of each galaxy in the $\log({\rm EW}_{{\rm H}\alpha})$ vs $g-r_t$ parameter space, and we then assign a slope value to this point using the IDWI method.}. Although we also considered alternative representative measures, such as the mean and median of all resolved slopes, our results using these alternatives did not show significant differences. However, $\bar{\alpha}$ is preferred because it reflects the central tendency of the resolved IMF-slope distribution. This quantity provides a robust indicator of the typical IMF-slope value across different regions of the galaxy, whereas the mean and median can be influenced by outliers or skewed distributions.

The comparison between the integrated IMF-slope ($\alpha_{int}$) and the resolved IMF-slope ($\bar{\alpha}$) is illustrated in Figure~\ref{res_vs_int}. A notable observation is that many data points cluster close to the one-to-one line (indicated by the black dotted line). For most galaxies, $\bar{\alpha}$ tends to be flatter than the integrated slope. However, when the integrated slope is flatter than approximately $\alpha_{int}>-2.7$, the relationship between the two slopes becomes closer to the one-to-one line.

This observation is further illustrated in Figure~\ref{int_res_hist}. We find that 71$\%$ of the galaxies have a flatter $\bar{\alpha}$ than their integrated slope ($\alpha_{int} - \bar{\alpha} < 0$), while 29$\%$ of the galaxies show the opposite trend, with a flatter $\bar{\alpha}$. Additionally, approximately 93$\%$ of the galaxies have $|\alpha_{int}-\bar{\alpha}|<0.5$, and $|\alpha_{int}-\bar{\alpha}|<0.25$ for about 81$\%$.
A difference of this magnitude is comparable to our typical uncertainties in the estimate of $\alpha_{res}$ ($\Delta \alpha_{res} \sim 0.167$), and so is unlikely to be significant. This only indicates that the distribution of $\alpha_{res}$ closely aligns with the value of $\alpha_{int}$, rather than implying that there is no variation in the slope within galaxies. This comparison validates our approach using resolved data, allowing us to effectively compare our findings with previous results where spatially resolved data is unavailable.

We also differentiate, in Figure~\ref{res_vs_int}, between galaxies with central star formation (represented by pink circles) and those without it (represented by black crosses). Galaxies with central star formation are defined as those in which the central $0.5 R_e$ region contains spaxels with a S/N at 4100\AA\ $>$ 3pix$^{-1}$, and at least 10$\%$ of these spaxels classified as star-forming or intermediate (those where emission likely results from a combination of star-forming and non-star forming mechanisms). Spaxel classification is based on the detection of specific emission lines, using either the [NII] (6583\AA\ )/H$\alpha$ versus [OIII] (5007 \AA\ )/H$\beta$ BPT diagram, or just the [NII] (6583\AA\ )/H$\alpha$ ratio. The spectral classification methodology for spaxels and galaxies is described in detail in \cite{owers_2019}.

In our sample, 182 objects exhibit no central star formation, while 1,162 objects do. Most galaxies without central star formation sit outside the bulk of the data points. For these galaxies, the integrated light is likely dominated by emission from older stellar populations rather than recent star formation. Consequently, the Kennicutt method may be less reliable in these cases, potentially leading to the inferred steeper IMF slope.

We investigate the resolved IMF slope distributions in greater detail for six galaxies in Figure~\ref{resolved_slope_hist}. These galaxies are randomly selected as a representative collection to illustrate the distributions of $\alpha_{res}$ within galaxies. In this figure, $\alpha_{int}$ is represented by a coloured vertical line, while the Salpeter slope is indicated by a black line for reference. Each galaxy exhibits a distinct range of slope distributions, with some showing broader or narrower slope ranges. Notably, the peak of the slope distribution does not necessarily align with either the Salpeter slope or the integrated slope.

To understand why the breadth of slope distributions varies among galaxies, we analyse the relationship between the IMF slope and other resolved properties of the galaxies in the next section. If these properties are related to the IMF slope, they could help explain the observed differences in the shapes of the slope distributions.

\subsection{Relation between $\alpha_{res}$ and other resolved properties}
\label{subsec:resolved_SFRD}

Given the direct link between the IMF slope and the star formation process, it is reasonable to anticipate that it may be associated with parameters such as the SFR and the $\Sigma_{\rm SFR}$. This section aims to quantify the relationship between these quantities more systematically, looking for an IMF slope driver.

In Figure~\ref{slope_vs_SFR} and Figure~\ref{slope_vs_SFRD}, we analyse the relationship between $\alpha_{int}$ and the global SFR, as well as $\alpha_{res}$ and $\alpha_{int}$ with the resolved and global $\Sigma_{\rm SFR}$, respectively. We do not compare $\alpha_{res}$ and the SFR per spaxel as this is fundamentally giving the same information than $\Sigma_{\rm SFR}$. To facilitate this analysis, we divide our sample into bins based on SFR and $\Sigma_{\rm SFR}$. Bins span $\log(\rm{SFR})=[-5.5,-1.5]$ and $\log(\Sigma_{\rm{SFR}})=[-4.5,-0.5]$ in steps of $0.5$. For each bin, we determine a representative IMF slope ($\bar{\alpha}_{res}$ and $\bar{\alpha}_{int}$) by identifying the maximum density coordinates of all spaxels (for resolved IMF slopes) or galaxies (for integrated IMF slopes) within that bin. The IMF slope at each coordinate is then assigned using the IDWI criterion, as outlined in \S\ \ref{sec:methods}.

It is evident that the IMF slope exhibits a strong correlation with both the SFR and $\Sigma_{\rm SFR}$, with a trend toward flatter $\bar{\alpha}_{int}$ at higher values of SFR and $\Sigma_{\rm SFR}$. In Figure~\ref{slope_vs_SFR}, we additionally examined the IMF slope as a function of SFR across three stellar mass bins: $M < 10^9 M_\odot$, $10^9 M_\odot \leq M \leq 10^{10} M_\odot$, and $M > 10^{10} M_\odot$. Our analysis revealed that although all mass bins display a similar qualitative trend, lower-mass galaxies tend to have flatter slopes at a given SFR compared to mid-mass and high-mass galaxies.

When comparing our results to other published work, we find that our measurements fill the gap between the low-SFRs probed by \cite{weidnet_2013} and the higher SFRs from \cite{lee_2009} and \cite{gunawardhana_2011}. It is important to note that our integrated measurements are derived from a 1 $R_e$ aperture, whereas previous studies use a variety of photometric measurements that would correspond to a range of different effective apertures. This may introduce subtle systematics in this comparison, but given the challenges in any similar IMF analysis, these are unlikely to be a significant contributor.

As illustrated by the black line in Figure~\ref{slope_vs_SFRD}, there is a clear relationship between $\bar{\alpha}_{res}$ (panel \textit{a)}) and $\bar{\alpha}_{int}$ (panel \textit{b)}) with $\Sigma_{\rm SFR}$, where higher $\Sigma_{\rm SFR}$ values are associated with flatter IMF slopes. Similar to the slope-SFR relationship, the integrated IMF slope exhibits a mass dependency with respect to $\Sigma_{\rm SFR}$: high-mass galaxies tend to have steeper slopes compared to mid- and low-mass galaxies for a given $\Sigma_{\rm SFR}$. However, this mass dependency is not observed in the resolved IMF slopes with respect to $\Sigma_{\rm SFR}$. This suggests that $\Sigma_{\rm SFR}$ may be an underlying physical driver of the IMF. This points to an scenario where the IMF is more closely linked to local parameters such as $\Sigma_{\rm SFR}$ rather than global properties like galaxy mass. Further exploration of the cause of this relationship is discussed below in \S\,\ref{sec:discussion}.

By fitting a linear model to the relationship between resolved IMF slopes and $\Sigma_{\rm SFR}$ across all mass ranges (depicted by the solid pink line), we derive the following general equation:
\begin{equation}
    \alpha_{res} = 0.26 \log(\Sigma_{\rm{SFR}}) - 2.03 \pm 0.036, \label{equation_}
\end{equation}
where $\Sigma_{\rm{SFR}}$ is in units of $\rm{M}_\odot \, \rm{yr}^{-1}\rm{kpc}^{-2}$. To estimate the uncertainty in Equation \ref{equation_}, we employed a Jackknife resampling method. Specifically, we randomly removed 10$\%$ of the data points from each bin of $\Sigma_{\rm{SFR}}$ and recalculated the linear fit multiple times. We then use the standard deviation of slopes ($\sigma_A$) and intercepts ($\sigma_B$) from these repeated fits to estimate the total uncertainty of the linear fit as $\text{ERR}_{\alpha_{\rm{res}}}=\sqrt{[\log(\Sigma_{\rm{SFR}}) \cdot \sigma_A]^2+\sigma_B^2}$.

In Figure~\ref{slope_vs_SFRD_mass}, we further examine the distribution of individual spaxels in the $\alpha_{res}$-SFR space. As expected, panel \textit{b)} of Figure~\ref{slope_vs_SFRD_mass} shows that spaxels are clustered around the binned relation lines of $\bar{\alpha}_{res}$ with $\log(\rm{SFR})$, as depicted in Figure~\ref{slope_vs_SFRD}. Notably, a mass dependence emerges within the $\alpha_{res}$-SFR relationship. The steep $\alpha_{\rm res}$ values observed around $\log(\Sigma_{\rm SFR}) \sim -5$ likely arise from limitations in our slope-estimation methods and the coverage of our SPS models.

This figure underscores that the observed slope-SFR relation in previous figures is not merely an artefact of the binning method but reflects a genuine physical connection between these quantities. Spaxels follow a moderate correlation with SFR, with a Spearman correlation coefficient of $\rho_{\rm{SFR}}=0.415$ (p$\sim$0). Although these relation exhibit significant scatter, much of this variability is likely due to measurement uncertainties. For example, the colour error bar in Figure~\ref{all_data_tracks} spans several tracks, corresponding to $\Delta \alpha_{res} \sim 0.167$. Despite this, we focus more on the observed trends rather than the exact numerical values. The strong correlations depicted in Figures~\ref{slope_vs_SFR} and \ref{slope_vs_SFRD} suggest a general relationship between the IMF slope and both SFR and $\Sigma_{\rm SFR}$, applicable to both resolved and global slopes. 

\subsection{Radial dependency of $\alpha_{res}$}
\label{subsec:radial_alpha_res}

To investigate a potential radial dependency of $\alpha_{res}$, we calculated the average value of $\alpha_{res}$ for spaxels within each radial bin. Each bin is defined by a radius that encompasses 20$\%$, 40$\%$, 60$\%$, 80$\%$, and 100$\%$ of the spaxels. The results are presented in Figure \ref{radial_p}, which shows the radial profiles of $\alpha_{res}$ for galaxies in three distinct mass bins. For visualisation purposes, each panel displays profiles for a representative subset of 75 randomly selected galaxies. 

Using the $\log(\Sigma_{\rm{SFR}})$-radius ($R$) relations from \cite{medling_2018} for galaxies on the main sequence (see their Figure 7), combined with our derived relationship between $\log(\Sigma_{\rm{SFR}})$ and $\alpha_{res}$ (equation \ref{equation_}), we predict the relationship between radius and $\alpha_{res}$. This predicted relation is the weak radial dependence shown as the red lines in Figure \ref{radial_p}. While some galaxies do exhibit variations in $\alpha_{res}$ with radius, the majority of objects have relatively flat radial profiles with minimal or no variation. The predicted shape is only marginally consistent with our measured values of $\alpha_{res}$ for $R/R_e < 2$. Given the uncertainties in our measurements, though, this relation may still be consistent with what we find, but there is clearly no strong signal across the sample investigated here.

Additionally, Figure \ref{radial_p} includes the mean $\alpha_{\rm{res}}$ values calculated across six radial bins, ranging from $\rm{R}/\rm{R_e} = 0$ to $3$ in increments of $0.5$. These mean values align somewhat more closely with the predicted relation (red lines), suggesting that there may be a weak underlying radial trend in $\alpha_{\rm{res}}$ that is not immediately evident in individual profiles, although this is probably not quite as strong as predicted.

\begin{figure*}
\centering
\includegraphics[width=0.8\linewidth]{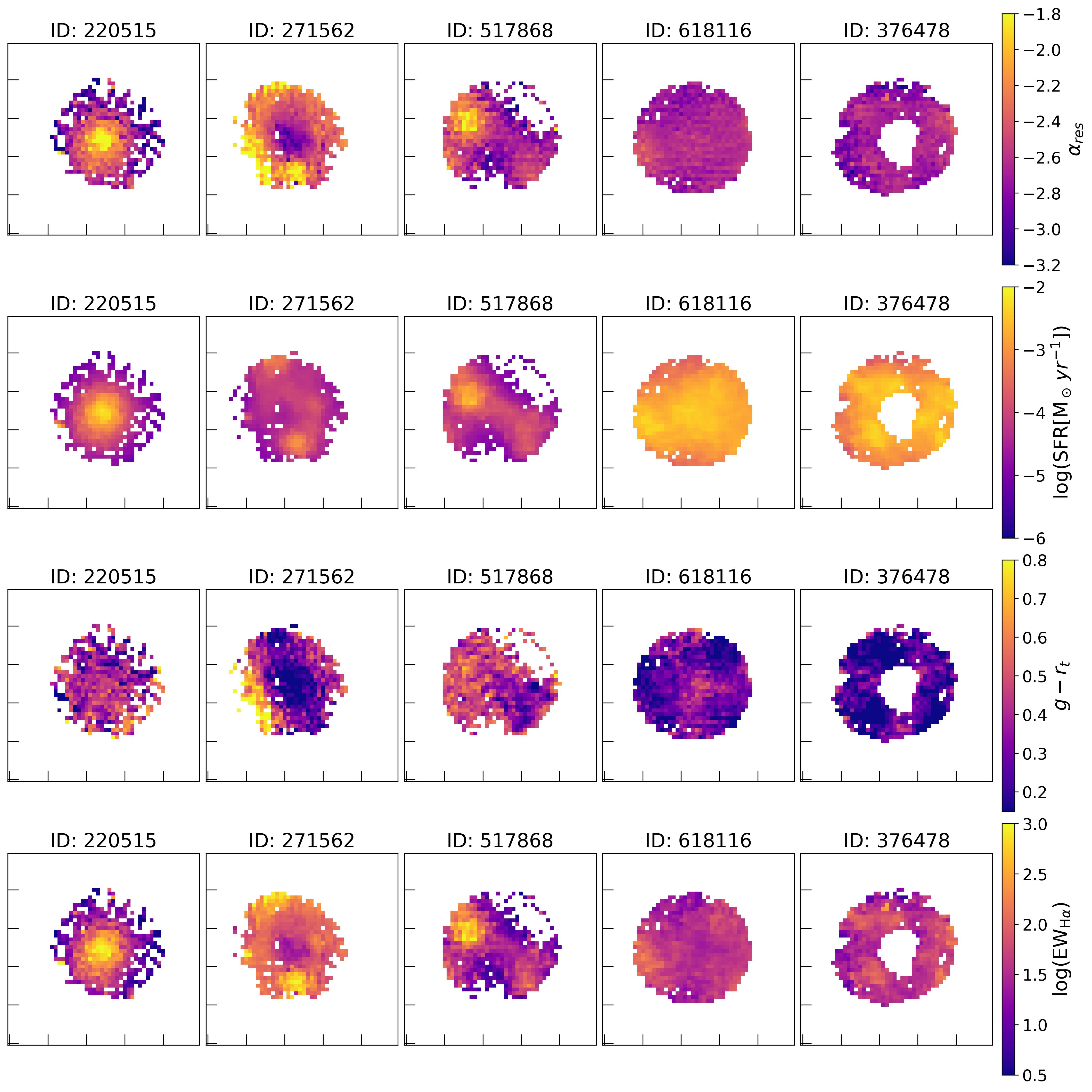}
\caption{This figure showcases maps of five example galaxies, each represented by a column. From top to bottom, the rows display the following quantities: $\alpha_{\rm{res}}$, $\log(\rm{SFR})$, $g - r_t$ colour, and $\log(\text{EW}_{\text{H}\alpha})$. These galaxies were selected to highlight diverse spatial features. Galaxies 220515 and 271562 exhibit clear radial gradients, while 517868 shows a non-central region with steeper IMF slopes. In contrast, galaxy 618116 has a relatively uniform (flat) distribution. Finally, galaxy 376478, which lacks central star formation, displays a range of steep IMF slopes. These examples illustrate the structural diversity and star formation characteristics present in the sample.}
\label{IMF_map_9}
\end{figure*}

\begin{figure*}
\centering
\includegraphics[width=0.8\linewidth]{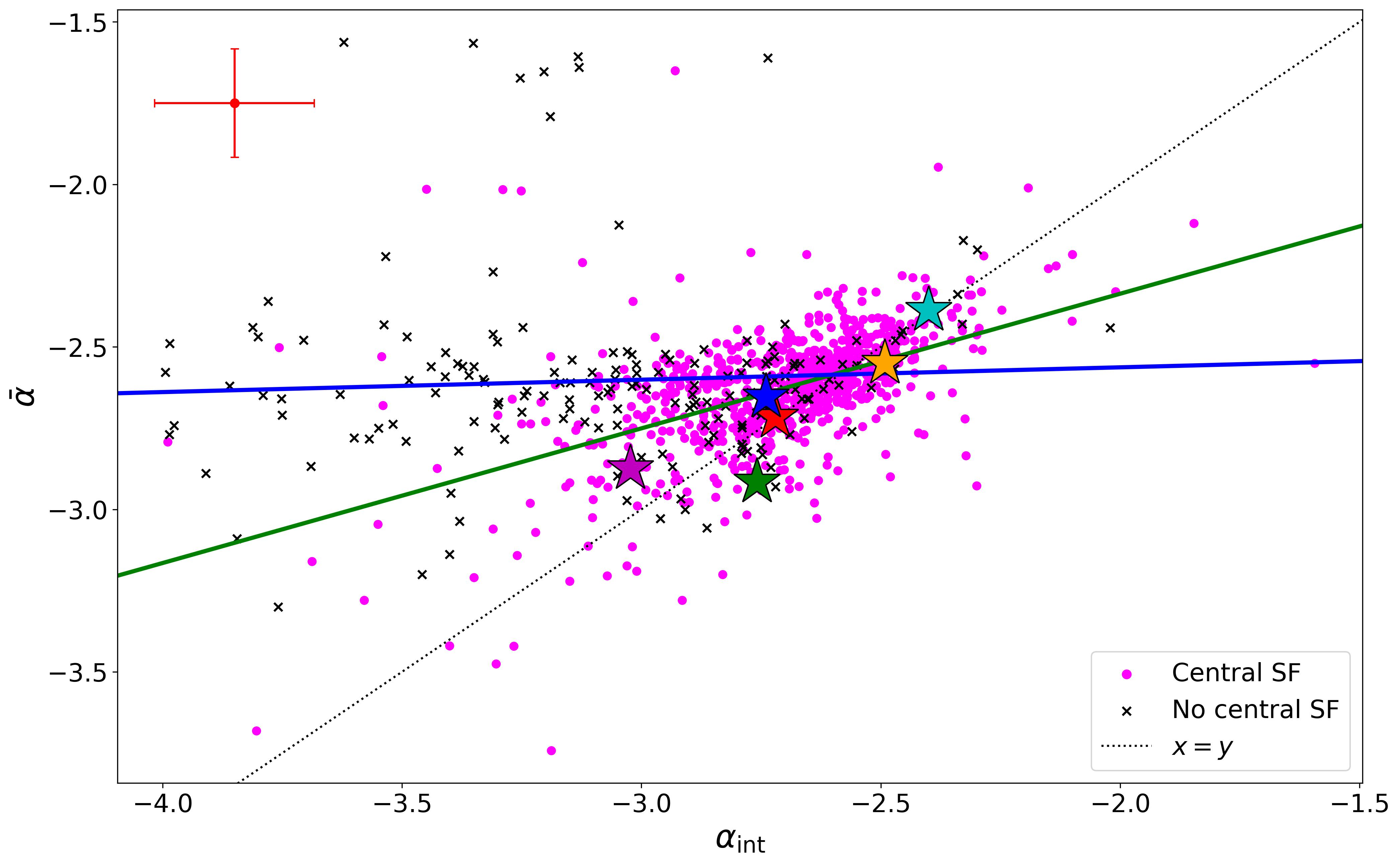}
\caption{This figure illustrates the relationship between the high-mass IMF slope ($\bar{\alpha}$) and the integrated IMF slope ($\alpha_{\rm{int}}$), focusing on galaxies where at least 50$\%$ of the spaxels are classified as star-forming. Pink circles represent galaxies with central star formation, while black crosses indicate galaxies without central star formation. The black dotted line denotes the one-to-one relationship, serving as a reference for comparison. Additionally, the green and blue lines show the best linear fits for galaxies with and without central star formation, respectively. The red cross in the corner indicates the typical uncertainties in the IMF. This plot highlights the differences between local and global IMF slopes.}
\label{res_vs_int}
\end{figure*}

\begin{figure}
\centering
\includegraphics[width=0.8\linewidth]{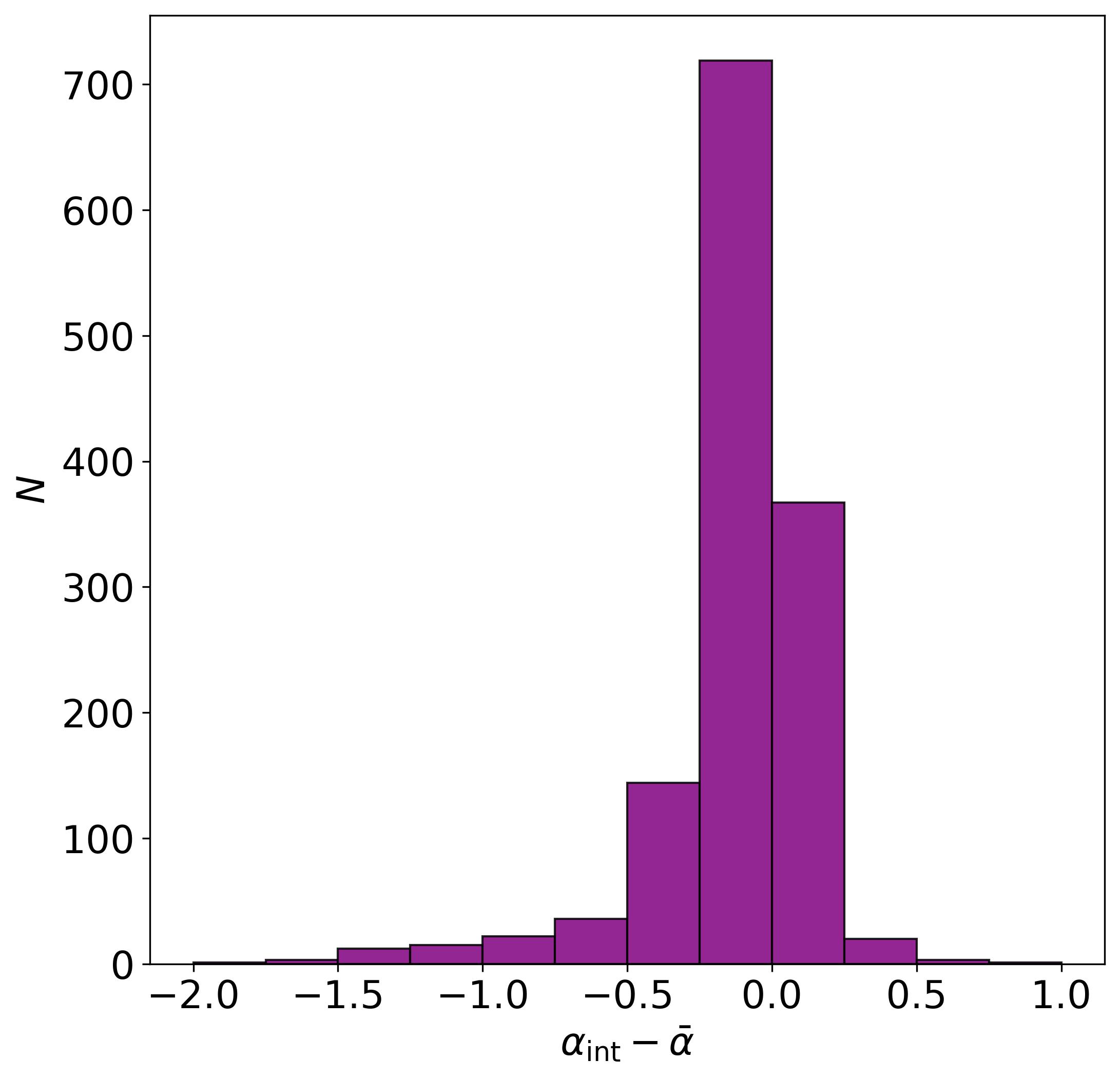}
\caption{The distribution of differences between the integrated IMF slope ($\alpha_{int}$) and the $\bar{\alpha}$ across our sample.}
\label{int_res_hist}
\end{figure}

\begin{figure*}
\centering
\includegraphics[width=0.8\linewidth]{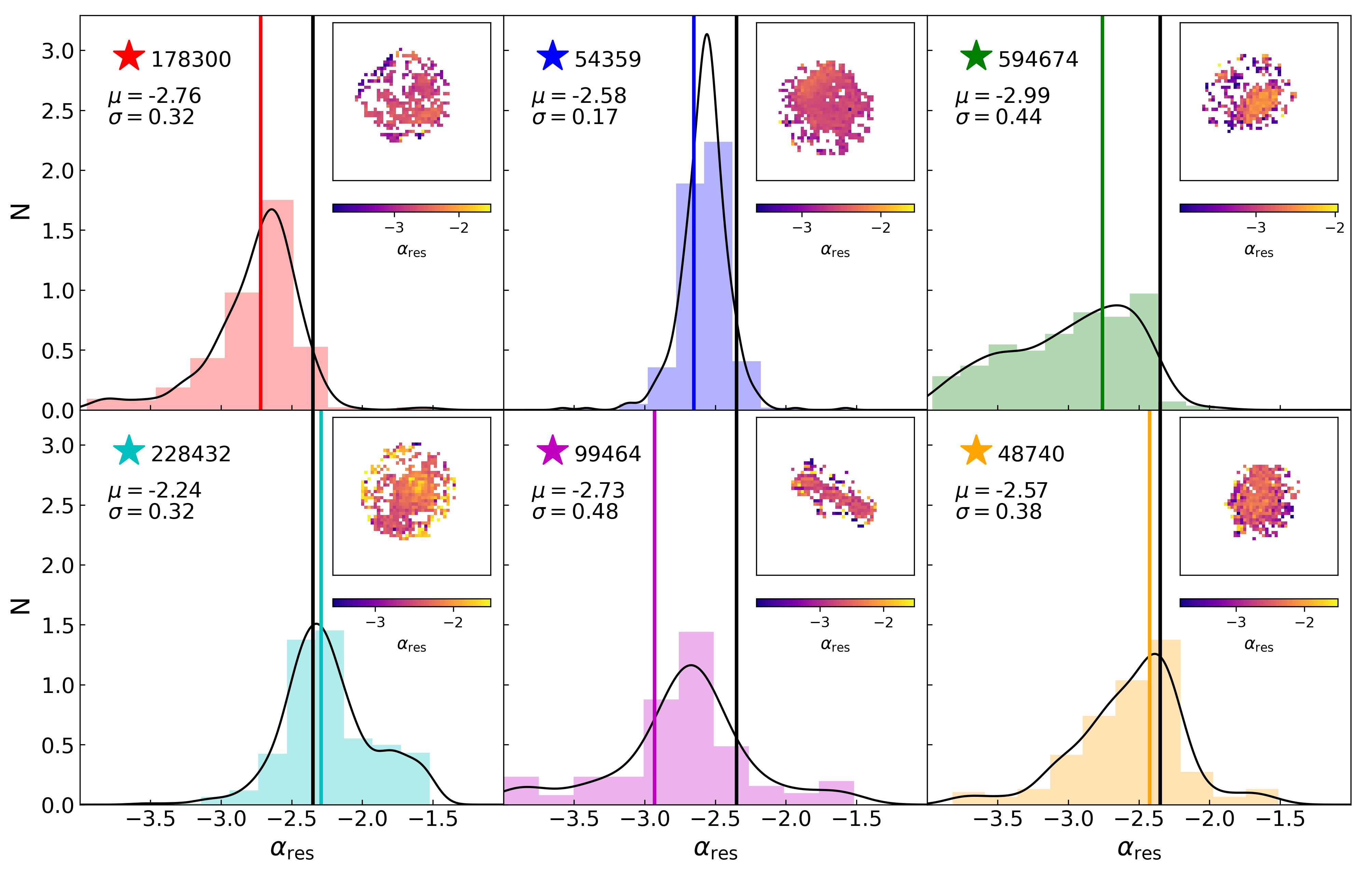}
\caption{This figure displays the distribution of the high-mass IMF resolved slope ($\alpha_{\rm{res}}$) for the six coloured galaxies presented in Figure~\ref{all_data_tracks_coloured}. Each distribution is represented by a kernel density estimation (KDE) shown as a black line. The vertical black line indicates the Salpeter slope ($\alpha = -2.35$), while the vertical coloured lines represent the integrated slopes for each galaxy. In the top left corner of each panel, the mean and standard deviation of the slope distribution are provided. Additionally, the $\alpha_{\rm{res}}$ map for each galaxy is displayed in the top right corner, offering a visual context for the slope distributions.}
\label{resolved_slope_hist}
\end{figure*}

\begin{figure*}
\centering
\includegraphics[width=0.8\linewidth]{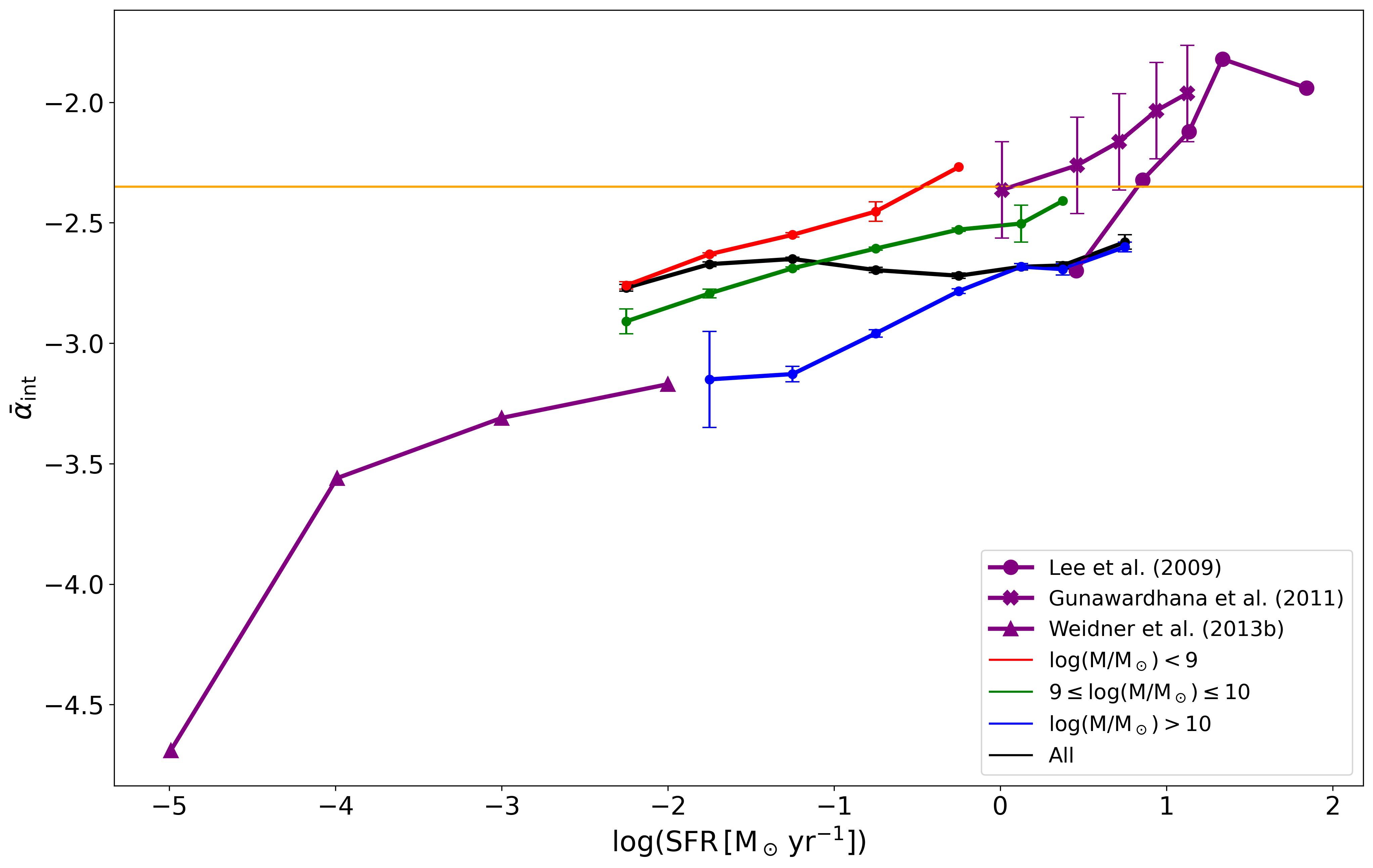}
\caption{This figure illustrates the relationship between the integrated high-mass IMF slope ($\bar{\alpha}{int}$) and the logarithm of the star formation rate ($\log(\rm{SFR})$) across all galaxies in the sample. The colour of each line denotes different stellar mass bins, highlighting the variations in slope with respect to star formation activity. Uncertainties are determined using a Jackknife resampling method. Additionally, results from previous studies by \cite{lee_2009}, \cite{gunawardhana_2011}, and \cite{weidnet_2013} are represented in purple for comparison. The Salpeter slope is shown as an orange line for reference.}
\label{slope_vs_SFR}
\end{figure*}

\begin{figure*}
\centering
\includegraphics[width=0.8\linewidth]{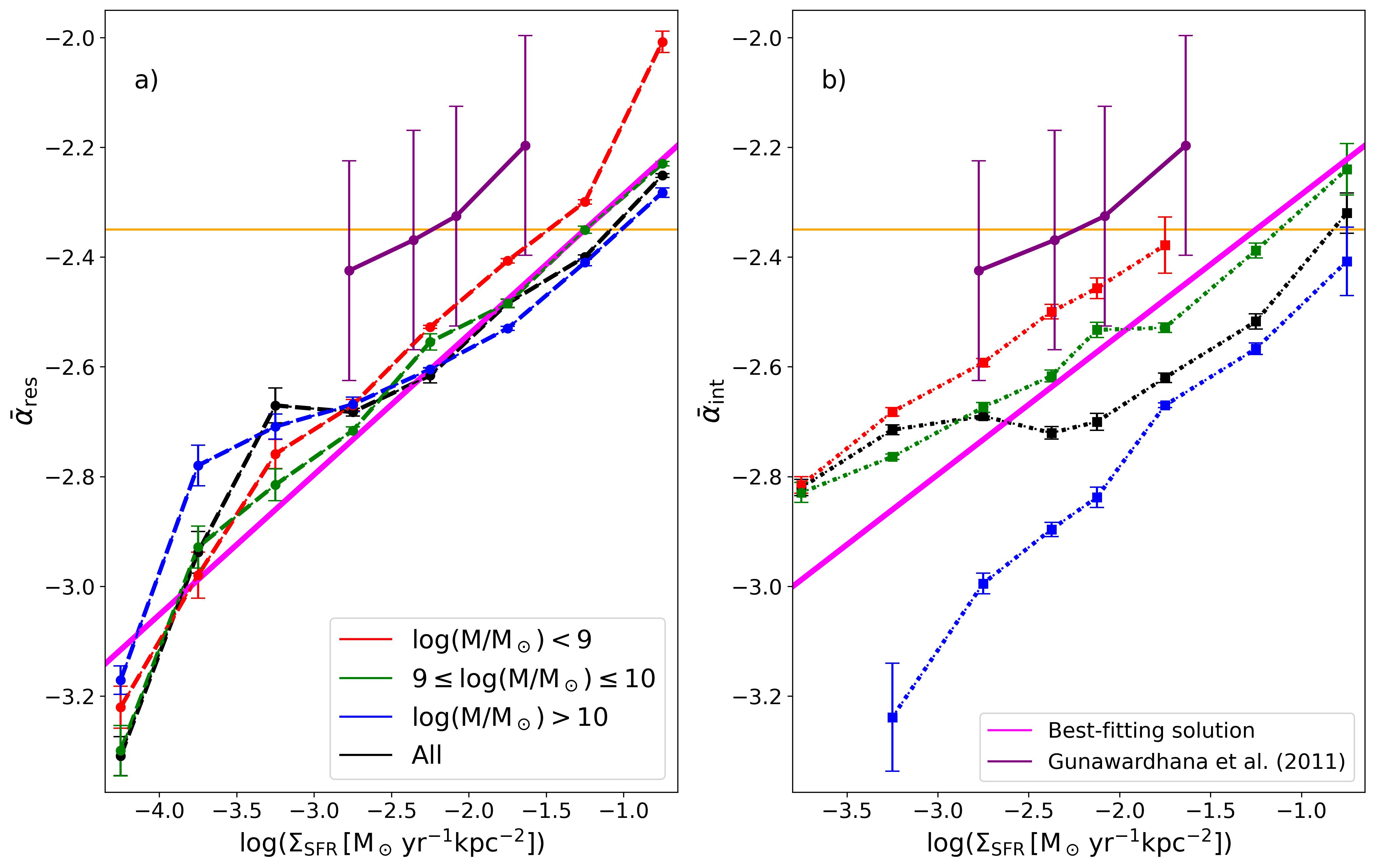}
\caption{This figure depicts the relationship between the resolved high-mass IMF slope (panel \textit{a}) and the integrated high-mass IMF slope (panel \textit{b}) against $\log(\Sigma_{\rm SFR})$. We make use of eight uniform bins of $\Sigma_{\rm SFR}$, associating an IMF slope to the coordinate of maximum density within the distribution of spaxels/galaxies in each bin. The colour of each line corresponds to different stellar mass bins. Uncertainties in the relationships are derived through Jackknife resampling resampling within each bin. The pink line in both panels represents the best linear fit for the relationship between the resolved $\bar{\alpha}$ and $\log(\Sigma_{\rm SFR})$ considering all stellar masses. The orange line indicates the Salpeter slope for reference.}
\label{slope_vs_SFRD}
\end{figure*}

\begin{figure*}
\centering
\includegraphics[width=0.9\linewidth]{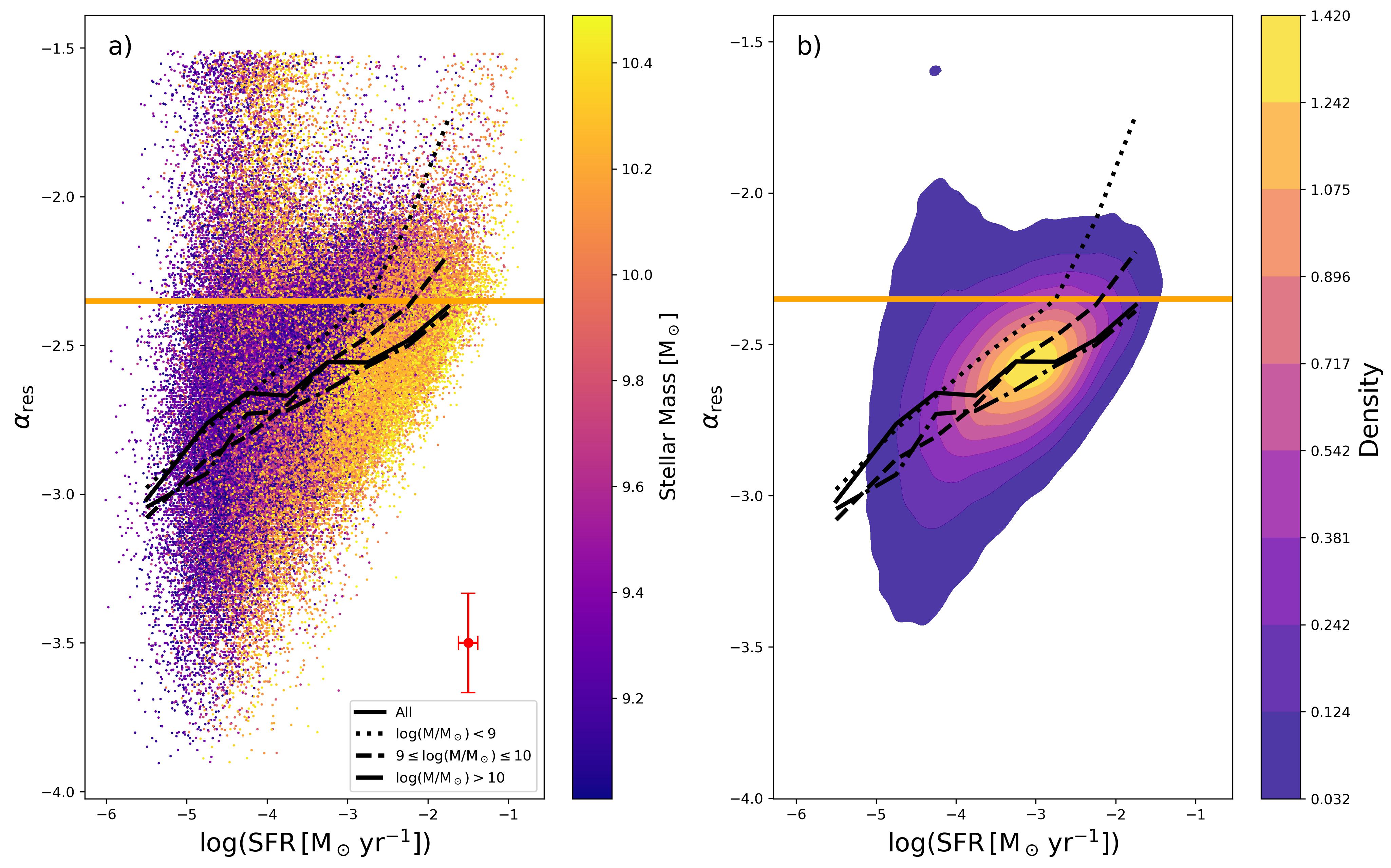}
\caption{Panel \textit{a)} displays the resolved high-mass IMF slope for each spaxel as a function of $\log(\rm{SFR})$. The black lines represent the $\bar{\alpha}_{res}$-$\log(\Sigma_{SFR})$ relations derived from Figure~\ref{slope_vs_SFRD}. The red cross in the bottom right corner indicates the uncertainties along both axes. Panel \textit{b)} illustrates the contours of the spaxel distributions from panel \textit{a}. It is evident that the majority of the data points align closely with the black lines. The steep $\alpha_{\rm res}$ values observed around $\log(\Sigma_{\rm SFR}) \sim -5$ likely arise from limitations in our slope-estimation methods and the coverage of our SPS models. For reference, the orange line indicates the Salpeter slope.}
\label{slope_vs_SFRD_mass}
\end{figure*}

\begin{figure*}
\centering
\includegraphics[width=0.9\linewidth]{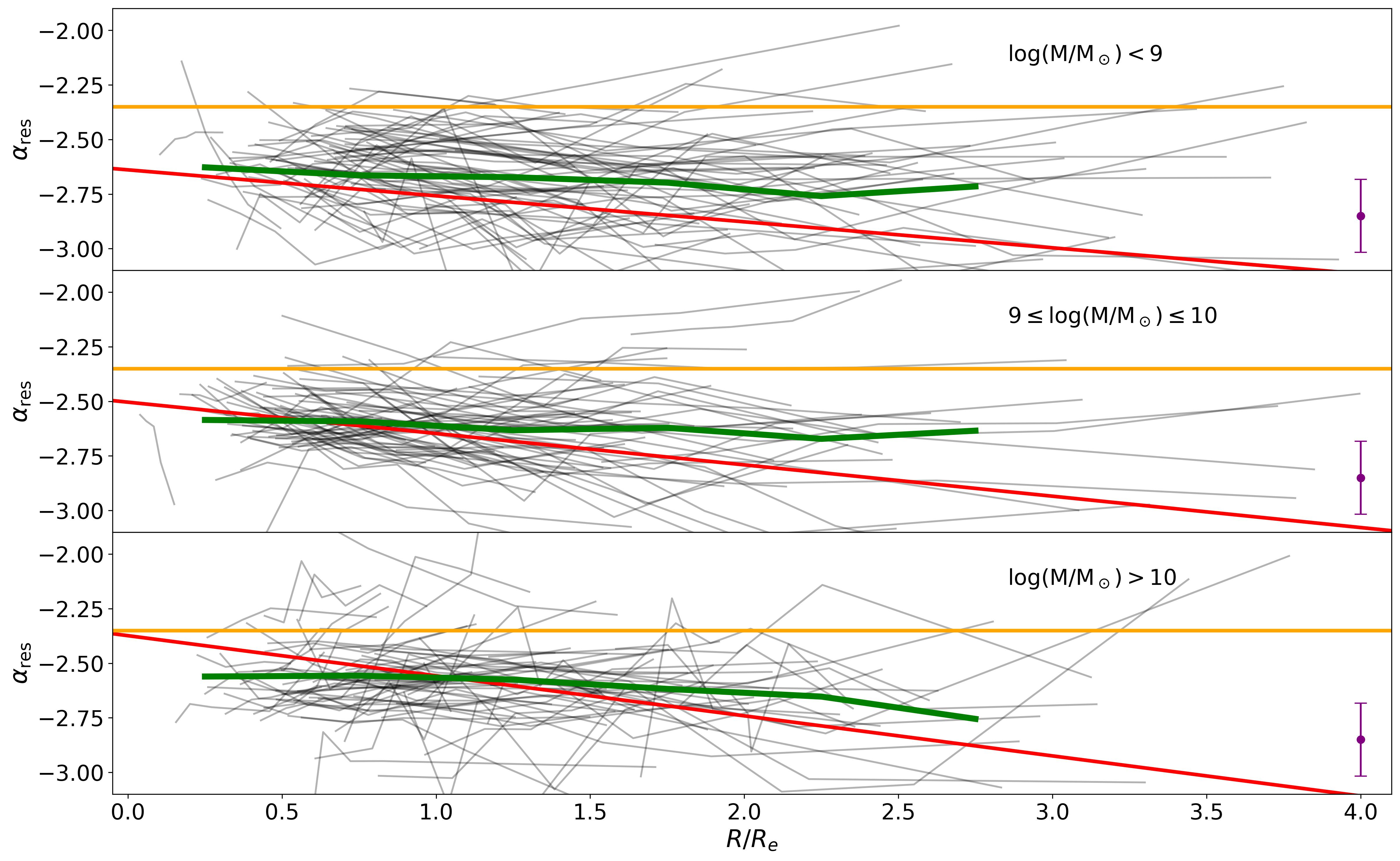}
\caption{Radial profiles of the resolved high-mass IMF slope ($\alpha_{\rm{res}}$) for 75 randomly selected galaxies across different mass bins. In each panel, the black lines represent the average $\alpha_{\rm{res}}$ values calculated for spaxels within defined radial bins. These radial bins are determined by radii that encompass annuli spanning 20$\%$, 40$\%$, 60$\%$, 80$\%$, and 100$\%$ of the total spaxels in each galaxy. The green lines represent the mean $\alpha_{\rm{res}}$ values of the radial profiles across 6 radial bins, ranging from $\rm{R}/\rm{R_e} =$ 0 to 3 in increments of 0.5. The red lines indicates the predicted relationship between radius ($R$) and $\alpha_{\rm{res}}$, derived from the $\Sigma_{\rm{SFR}}$-$\rm{R}$ relation of \cite{medling_2018} and our $\Sigma_{\rm{SFR}}$-$\alpha_{\rm{res}}$ relation (as outlined in equation \ref{equation_}). The purple cross in the corner represents the uncertainties in the IMF. For reference, the orange line denotes the Salpeter slope.}
\label{radial_p}
\end{figure*}

\section{DISCUSSION}
\label{sec:discussion}

\subsection{Limitations on the method}

Given the innovative aspects of our approach, it is crucial to recognise the inherent limitations and potential sources of uncertainty within our methodology.

The accuracy of our results is highly dependent on the capabilities of the SPS tool employed. We opted to use the \texttt{P{\'E}GASE} SPS tool primarily due to its flexibility in modifying the IMF functional form, adjusting lower and upper mass ranges, SFHs, and slopes. However, \texttt{P{\'E}GASE} does not account for certain stellar properties, such as binary interactions and rotation \citep{leitherer_1999,stanway_2016, eldridge_2012,eldridge_2017}, which could limit its accuracy in comparison to more complete SPS tools and may introduce some degree of uncertainty in our IMF estimation. 

As discussed by \cite{conroy_2013}, systematic uncertainties in SPS models can lead to variations in IMF estimates. Nevertheless, previous studies by \cite{gunawardhana_2011} and \cite{nanayakkara_2017,nanayakkara_2020} have demonstrated that results obtained with \texttt{P{\'E}GASE} are not significantly different from those generated by other SPS tools, such as \texttt{STARBURST99} \citep{leitherer_1999}. The underlying limitation remains that spectral synthesis models are constrained by the accuracy of the stellar models they incorporate \citep{hoversten_2010}. Therefore, while the specific quantitative results may vary when using different population synthesis tools, we expect the qualitative trends observed in our work to remain robust.

Given that this work is among the first to apply this methodology for estimating the IMF on a spaxel-by-spaxel basis, it is still too early to fully understand whether the methods for estimating the IMF encounter additional limitations when applied at more resolved scales. Our IMF estimation method may require a higher signal-to-noise ratio (S/N) on smaller scales to retrieve reliable results. One advantage of conducting the analysis with integrated light is that the effects of different stellar populations are smoothed out, as noted by \citet{hoversten_2008}. However, a more localised approach may require special attention to the distinct stellar populations present. As a galaxy evolves, the stellar population present in any given spaxel is likely to consist of a mix of populations that formed with different ages, metallicities, and potentially IMF shapes. This could significantly impact the accuracy of our estimation of $\alpha_{res}$. Conversely, the similar results that we see, broadly speaking, between $\alpha_{res}$ and $\alpha_{int}$ are reassuring, suggesting that this effect is not introducing a significant bias or systematic. Also, in this analysis we estimate the high mass IMF slope, which is dominated by short lived high mass stars. The lifetime of such stars is negligible compared to the rotation speed of galaxies that would drive population mixing. This strongly mitigates against any biases arising through the possibility of population mixing, which may be a more substantial concern when analysing the low mass, long-lived, stellar populations. Further studies are needed to explore these effects in greater detail.

Another limitation of our IMF estimates lies in the method we use to assign a slope to each spaxel or galaxy based on its location in the $\log({\rm EW}_{{\rm H}\alpha})$ vs $g-r_t$ space. Although we have a range of spectral evolution models that encompass the majority of the objects of interest, a certain percentage falls outside the region covered by these models. For those spaxels, our interpolation model becomes unreliable, as extrapolation would be required. However, extrapolated points may deviate significantly from what theoretical models would predict. It is possible to generate additional models to cover the portions of the $\log({\rm EW}_{{\rm H}\alpha})$ vs $g-r_t$ space that are not currently spanned by including different parameters, such as varying SFHs and metallicities. Nevertheless, as mentioned above, this issue affects only a small fraction of our data (less than $8\%$ of all spaxels) and does not significantly impact our overall conclusions. In a similar vein, \cite{nanayakkara_2020} demonstrated that incorporating stochastic starburst events into the SFHs of galaxies can lead to higher $\rm{EW}{\rm{H}\alpha}$ values. This would result in a modified distribution of the models in the $\log({\rm EW}_{{\rm H}\alpha})$ versus $g-r_t$ space. This suggests that some portion of the variation in the inferred IMF slopes seen in our results may actually stem from differences in SFHs, metallicities or other properties. Incorporating the full suite of SFHs and metallicities as additional parameters would significantly increase the complexity of any investigation, and is beyond the scope of this current analysis.

Finally, we note that we compared the resolved slope of star-forming spaxels with the slope obtained from the integrated light of galaxies, which includes both star-forming and non-star-forming spaxels, and thus different stellar populations. In future work, we aim to explore methods to estimate the slope in these non-star-forming spaxels to construct complete IMF slope maps and make more robust comparisons.

\subsection{Understanding the link between $\alpha_{res}$ and $\alpha_{int}$}

The IMF has often been assumed to be universal, typically adopting the Salpeter slope ($\alpha = -2.35$), likely because it provides a simple approach and reasonably accurate assumption. However, there is an overwhelming body of evidence suggesting the necessity of considering variable IMFs.

These findings suggest that the IMF is influenced by various properties, driving its variability \citep{navarro_2019,barbosa_2021,navarro_2021,navarro_2023}. Given that star formation conditions within a galaxy can vary significantly (e.g., different metallicities, magnetic field intensities, turbulence), a critical question arises: at what scale does the IMF begin to vary? Specifically, do the local variations in the IMF (e.g., at the scale of molecular clouds or star clusters) average out when considering the global scale (e.g., an entire galaxy), or does the global IMF robustly depend on local variations, leading to IMF differences between galaxies? The more fundamental question is whether there is a (continuous) connection between the local and global IMF.

The qualitative consistency observed between $\bar{\alpha}_{res}$ and $\bar{\alpha}_{int}$ in their relationships with SFR and $\Sigma_{\rm SFR}$ (Figures \ref{slope_vs_SFR} and \ref{slope_vs_SFRD}) suggests a fundamental connection between the local and global IMF properties within galaxies. While the explicit nature of this relationship requires further exploration, our findings indicate that local variations in the IMF contribute to the formation of the global IMF signature.

As illustrated in Figure~\ref{int_res_hist}, the difference between $\bar{\alpha}$ and $\alpha_{int}$ in our sample follows a distribution characterised by a mean of $\mu = -0.124$ and a standard deviation of $\sigma = 0.27$. The typical standard deviation of the resolved slope within a galaxy is approximately $0.34$. It is important to note that galaxies lacking central star formation generally exhibit lower values of $\alpha_{int}$ ($\alpha_{int}<-2.7$) while maintaining consistent $\bar{\alpha}$ values (Figure~\ref{res_vs_int}). This phenomenon is likely attributable to the influence of central pixels on the $\alpha_{int}$ value. These non-star-forming central pixels are likely contributing light from older stellar populations to the overall galaxy light. Given that the Kennicutt method relies on the strength of the H$\alpha$ line and the $g-r_t$ colour, this older stellar population may introduce bias into our IMF slope estimation. Supporting this explanation is the observation that, in galaxies with central star formation, $\bar{\alpha}$ and $\alpha_{int}$ tend to align more closely with the one-to-one relation.

It is important to acknowledge the assumptions made to facilitate the comparison between the resolved and integrated IMF slopes. First, we represent all the resolved slopes by a single value, $\bar{\alpha}$, for each galaxy, which does not represent a direct comparison between the full distribution of $\alpha_{res}$ with $\alpha_{int}$. Future studies could explore more appropriate methods to address this limitation. Second, our analysis focuses solely on spaxels associated with star formation, excluding those without such activity. This introduces a potential bias when comparing the slopes inferred from this sample to those estimated from the integrated galaxy light, which includes contributions from older stellar populations and may bias our IMF estimation less reliable. To mitigate this bias, we ensured that all galaxies in our sample have over 50$\%$ of their spaxels classified as star forming.

Despite all these limitations, we still find a reasonable agreement between $\alpha_{res}$ and $\alpha_{int}$. Our results suggest that local IMF variations significantly contribute to the global IMF signature, highlighting the importance of considering variable IMFs in understanding star formation across different scales.

\subsection{The connection with SFR and $\Sigma_{\rm{SFR}}$}

One of our key findings is the relationship between the IMF slope (both resolved and integrated) and the SFR and $\Sigma_{\rm SFR}$, where galaxies with higher SFR and $\Sigma_{\rm SFR}$ exhibit flatter IMF slopes. Since the IMF is directly linked to the star formation process, the observed relation between the IMF slope and these star formation indicators is likely associated with various factors critical to star formation. Star formation begins with the fragmentation and collapse of cold molecular clouds \citep{krumholz_2019}. Perturbed velocity fields and turbulence induce shocks, creating regions of compressed gas. As the density increases, gravitational collapse eventually overcomes turbulent support \citep{mckee_2007}. Consequently, the physical conditions of the ISM—such as turbulence, feedback, outflows, gas cooling/heating, magnetic fields, and metallicity—are expected to significantly impact star formation efficiency \citep{bate_2005,gouliermis_2018,grasha_2017}.

The Jeans mass of an unperturbed molecular cloud represents the threshold at which the system becomes gravitationally bound. In regions with a high star formation rate (SFR $> 3-5 M_\odot yr^{-1}$), radiation from hot stars can warm the surrounding dust, which then thermally couples with the gas. This process leads to an increase in the Jeans mass in high-density and high-temperature regions, as more mass is required to overcome the thermal and kinetic energy of the dust and gas, making star formation more difficult. Consequently, the SFR and Jeans mass are closely interconnected. As early as 1992, \cite{larson_1992,larson_1998,larson_2005} proposed a connection between the IMF and the Jeans mass. An increase in the Jeans mass results in fewer clumps forming within a star-forming cloud, but these clumps will concentrate more mass. This leads to a flatter high-mass IMF slope \citep{bonnell_2006}. This scenario helps explain why galaxies 618116 and 376478 in Figure~\ref{IMF_map_9} exhibit steep $\alpha_{res}$ values despite having high SFRs. When the Jeans mass is low, star formation becomes more efficient, resulting in the formation of numerous low-mass stars while inhibiting the formation of high-mass stars. This process consequently steepens the slope of the IMF.

\cite{bate_2005} conducted hydrodynamical simulations to model the formation of stellar systems from the collapse of molecular clouds. These outcomes were compared to earlier calculations \citep{bate_2003}, where the Jeans mass was a factor of three higher than the original value of 1 M$_\odot$. The comparison revealed that denser clouds tend to produce a higher proportion of brown dwarfs and exhibit higher velocity dispersions. The study found that the interaction between accretion and ejection processes could replicate low-mass IMFs, while variations in the magnitude of dispersion in the accretion rates of individual objects would primarily affect the slope of the high-mass IMF. Specifically, the high-mass IMF slope becomes flatter in denser star-forming regions. This finding aligns closely with our results, as well as those of \cite{gunawardhana_2011}. Additionally, \cite{bonnell_2006} determined that the ``knee'' of the IMF is approximately determined by the Jeans mass, and that a barotropic equation of state can produce a realistic IMF for high Jeans masses.

Hydrodynamical simulations by \cite{narayanan_2012} suggest that variations in the IMF, particularly through the scaling of Jeans mass with the IMF's characteristic mass, could potentially reconcile discrepancies between observed and model-predicted star formation rates. However, they acknowledge that these differences could also be addressed without assuming a varying IMF. Magnetic fields play a critical role in star formation, as shown by \cite{sharda_2020}, who found that strong magnetic fields can suppress gas cloud fragmentation, reducing the formation of low-mass stars. Despite this, other studies indicate that the impact of magnetic fields on star formation may not be as significant as radiative feedback \citep{bate_2011,myers_2013} and thermodynamic processes \cite{lee_2018a}.

We find that the high-mass slope of the IMF is primarily influenced by the SFR surface density, which may be explained by the formation of high mass stars altering the Jeans mass and fragmentation within molecular clouds. This is consistent with the Kennicutt-Schmidt law \citep{schmidt_1959,kennicutt_1989}, which posits a strong correlation between SFR and gas density. While it is beyond the scope of this work to determine which parameter or process predominantly suppresses star formation, evidence indicates that magnetic fields and radiative feedback have a non-negligible impact on star formation.

The observed stellar mass dependency in the slope-SFRD relation for integrated measurements, as shown in Figure~\ref{slope_vs_SFRD}, supports the idea that the IMF slope is determined by local processes. If the resolved IMF slopes depended on the galaxy's total stellar mass, we would need to explain how different regions across a galaxy adjust their IMF values to ensure alignment with such a global mass dependency trend. The presence of a mass dependency in the $\alpha_{int}$ vs $\Sigma_{\rm SFR}$ relation suggests that high-mass galaxies tend to have steeper IMF slopes, likely due to their older stellar populations, arising from local drivers. \cite{gunawardhana_2011} find a relationship between their $\alpha_{int}$ and specific star formation rate, which may also indicate an underlying local mass dependency.

There is a well-established relationship between $\Sigma_{\rm SFR}$ and galactocentric radius, as demonstrated by \cite{medling_2018}. In this work, we have identified a relationship between the IMF slope and $\Sigma_{\rm SFR}$, leading us to anticipate a corresponding relationship between the IMF slope and galactocentric radius. Indeed, studies by \cite{navarro_2015,navarro_2019}, \cite{davis_2017}, \cite{vaughan_2018}, and \cite{vandokkum_2017} have shown that massive early-type galaxies exhibit a radial dependency of the IMF, with $\alpha_{res}$ becoming steeper towards the outskirts of the galaxy. Our results, though, suggest a lack of significant radial dependencies in the spatial distribution of the IMF slope. We believe this is due to the high scatter in the $\alpha_{res}$ vs $\Sigma_{\rm SFR}$ relation (solid black line in panel a) in Figure \ref{slope_vs_SFRD_mass}), which likely weakens any radial dependencies. Further work is needed to more precisely quantify the spatial variations of the IMF slope. Recent work by \cite{navarro_2023}, which presents a low-mass IMF slope map for a star-forming galaxy, also indicates minimal radial dependency of the IMF slope. Radial IMF gradients have only been observed in quiescent ETGs so far, and only probe the low-mass end of the IMF, a regime distinct from the high-mass slope we focus on here. Further work is needed to verify the existence of radial gradient of the IMF high-mass slope of star forming galaxies. Given the precision of our $\alpha_{res}$ estimations, such gradients would only be detectable if they are sufficiently pronounced to overcome the limitations of our current accuracy.
 
\section{SUMMARY}
\label{sec:conclusion}

We have conducted the most extensive systematic analysis to date on the distribution of the high-mass IMF slope with spatially resolved spectroscopic galaxy data. We have used data from star-forming galaxies taken from the SAMI survey. The high-mass IMF slope is estimated using the Kennicutt method. To achieve this, we employ the \texttt{P{\'E}GASE} population synthesis tool to generate several galaxy evolutionary tracks, each corresponding to a different IMF slope. We then use a spatial interpolation technique (IDWI) to assign a slope to spaxels and galaxies based on their location in the $\log({\rm EW}_{{\rm H}\alpha})$ vs $g-r_t$ colour space.

Our study examines the relationship between the spaxel-scale resolved IMF slope and the global integrated galaxy light IMF slope, as well as the variations in the resolved IMF slope across different galaxies. We aim to explain these variations by exploring the connection between the IMF slope and galaxy properties such as SFR and $\Sigma_{\rm SFR}$. Our findings indicate that:
\begin{itemize}
    \item The IMF of local regions within a galaxy are largely consistent with the global IMF of the galaxy, particularly for galaxies with $\alpha_{int}\gtrsim-2.7$. However, the steepest IMF slopes derived from global measurements may be overestimated due to the contribution of spaxels with flatter IMF slopes in the inner regions of galaxies, particularly in those without central star formation.

    \item We attribute the higher difference between $\bar{\alpha}$ and $\alpha_{int}$ for steeper global IMF slopes to the reliability of the Kennicutt method, which depends on strong H$\alpha$ emission. Galaxies with higher star formation rates tend to have both a flatter IMF slope and stronger H$\alpha$ lines, thereby enhancing the accuracy of IMF slope estimation. In contrast, galaxies with lower star formation rates may lack strong H$\alpha$ lines, leading to greater uncertainty in our IMF estimates. 
    
    \item We find a variety of values of $\alpha_{res}$ within galaxies, (Figure~\ref{resolved_slope_hist}). While $\bar{\alpha}$ is typically not significantly different from the integrated IMF, the resolved IMF slope distribution can be wide, narrow, or even bimodal. Some galaxies have a radial gradient with $\alpha_{res}$, although the population as a whole does not show a strong radial trend. 

    \item We find a strong correlation between the resolved and integrated IMF slopes with SFR and $\Sigma_{\rm SFR}$ (Figure~\ref{slope_vs_SFRD}). Specifically, a flatter/steeper IMF slope is associated with higher/lower SFR and $\Sigma_{\rm SFR}$, both for local and global scales. We introduce Equation \ref{equation_} to link $\alpha_{res}$ to $\Sigma_{\rm SFR}$. 
    
    \item The $\bar{\alpha}_{int}$-SFR relation (Figure~\ref{slope_vs_SFR}) exhibit a mass dependency, where lower stellar mass galaxies tend to have shallower slopes at a given SFR. This mass dependency is also evident in the relation between $\bar{\alpha}_{int}$ and $\Sigma_{\rm SFR}$ (Figure~\ref{slope_vs_SFRD}), but it goes away when considering $\bar{\alpha}_{res}$. We suggest that regions with higher star formation density, in particular those characterised by more concentrated star formation, are likely to produce a larger number of high-mass stars, leading to a flattening of the high-mass IMF slope. This scenario would require an increase in the Jeans mass, which could be driven by factors such as turbulence, magnetic fields or other ISM processes.
\end{itemize}
The aim of this work is to present an analysis of the high-mass IMF slope on a spaxel-by-spaxel scale, with the goal of understanding the relationship between the IMF at this scale and the global IMF of galaxies. Additionally, we examine how the variability of the resolved slope correlates with other galaxy properties such as SFR and $\Sigma_{\rm SFR}$. Comparisons between recent JWST observations and semi-analytical models provide evidence supporting a top-heavy IMF at $z>10$ \citep{yung_2024}. These findings support our results, as the early universe features higher star formation density, and our study indicates a flatter high-mass IMF for in such environments. Future work will focus on enhancing the rigour of our approach by considering the SFH as a variable and exploring methods to estimate both the low-mass and high-mass resolved IMF slopes in regions with and without star formation. This could potentially allow us to further investigate differences in the distribution of the resolved slopes in passive and star forming galaxies across various environments. With this information, we can gain deeper insights into the origin of the IMF and its relationship not only with galaxy properties but also with their environment.

\begin{acknowledgement}
The SAMI Galaxy Survey is based on observations made at the Anglo-Australian Telescope. The Sydney-AAO Multi-object Integral field spectrograph (SAMI) was developed jointly by the University of Sydney and the Australian Astronomical Observatory. The SAMI input catalogue is based on data taken from the Sloan Digital Sky Survey, the GAMA Survey and the VST ATLAS Survey. The SAMI Galaxy Survey is supported by the Australian Research Council Centre of Excellence for All Sky Astrophysics in 3 Dimensions (ASTRO 3D), through project number CE170100013, the Australian Research Council Centre of Excellence for All-sky Astrophysics (CAASTRO), through project number CE110001020, and other participating institutions. The SAMI Galaxy Survey website is \url{http://sami-survey.org/}.
\end{acknowledgement}

\bibliography{bibtemplate}


\end{document}